\documentclass{aa}  

\usepackage{graphicx}
\usepackage{txfonts}
\usepackage{amsmath}
\usepackage{amssymb}
\usepackage{booktabs}
\usepackage{physics}
\usepackage{caption}
\usepackage{subcaption}
\usepackage{color,soul}
\def\eg{\emph{e.g. }}

\def\ie{\emph{i.e. }}

\begin{document}

   \title{Stellar mass and radius estimation using Artificial Intelligence}

   \author{A. Moya
          \inst{1,2}
          \and
          R. J. L\'opez-Sastre\inst{3}
          }

   \institute{Departament d’Astronomia i Astrofísica, Universitat de València, C. Dr. Moliner 50, 46100 Burjassot, Spain
\and
    Electrical Engineering, Electronics, Automation and Applied Physics Department, E.T.S.I.D.I, Polytechnic University of Madrid (UPM), Madrid 28012, Spain\\
              \email{andres.moya-bedon@uv.es}
         \and
             GRAM Research Group, Department of Signal Theory and Communications, University of Alcal\'a, Alcal\'a de Henares, 28805, Spain. \\
             \email{robertoj.lopez@uah.es}
             }

   \date{Received September 15, 1996; accepted March 16, 1997}

% \abstract{}{}{}{}{} 
% 5 {} token are mandatory
 
  \abstract
   {Estimating stellar masses and radii is a challenge for most of the stars but their knowledge is critical for many different astrophysical fields such as exoplanets characterization or stellar structure and evolution, for example. One of the most extended techniques for estimating these variables are the so-called empirical relations.}
   {In this work we propose a group of state-of-the-art AI regression models, with the aim of studying their proficiency in estimating stellar masses and radii. We select the one providing the best accuracy with the least possible bias. Some of these AI techniques do not properly treat uncertainties, but in the current context, where the statistical analysis of massive databases in different fields is being done, the most accurate estimation possible of stellar masses and radii can provide valuable information. We publicly release the database, the AI models, and an online tool for stellar mass and radius estimation to the community.}
   {We use a sample of 726 MS stars in the literature with accurate $M$, $R$, $T_{\rm eff}$, $L$, $\log g$, and [Fe/H]. We have split our data sample into training and testing sets and then analyzed the different AI techniques with them. In particular, we have experimentally evaluated the accuracy of the following models: Linear Regression, Bayesian Regression, Regression Trees, Random Forest, Support-Vector Regression (SVR), Neural Networks, kNN, and Stacking. We propose a series of experiments designed to evaluate the accuracy of the estimations, and also the generalization capability of AI models. We have also analyzed the impact of reducing the number of inputs parameters and compared our results with those from state-of-the-art empirical relations in the literature.}
   {We have found that a {\it Stacking} of several regression models is the most suitable technique for estimating masses and radii. In the case of the mass, Neural Networks also provide precise results, and for the radius, SVR and Neural Networks work too. When comparing with other state-of-the-art empirical relations based models, our {\it Stacking} improves the accuracy by a factor of two for both variables. In addition, bias is reduced to one order of magnitude in the case of the stellar mass. Finally, we have found that using our {\it Stacking} and only $T_{\rm eff}$ and $L$ as input features, the accuracies obtained are slightly larger than a 5$\%$, with a bias $\approx 1.5\%$. In the case of the mass, including [Fe/H] significantly improves the results. For the radius, it is the inclusion of $\log g$ that yields better results. Finally, the proposed AI models exhibit an interesting generalization capability, being able to perform estimations for masses and radii never observed during the training step.}
   {}


   \maketitle
%
%-------------------------------------------------------------------

\section{Introduction}

Among the different stellar parameters, stellar mass ($M$) and radius ($R$) are two of the variables with a larger impact for understanding different stellar physical processes. Unfortunately, they can only be indirectly estimated for most of the stars. In the literature, there are only two observational methods providing reliable estimations of stellar masses: Detached Eclipsing Binaries (EBs), and Asteroseismology (Ast). In the case of radii, we must add interferometry to the list. When applying these methods, almost all the isolated stars are out of their focus for physical or technical reasons. Nevertheless, in addition to these techniques, a large number of other techniques for estimating these parameters have been proposed/used \citep[see][for a complete review for the case of stellar masses]{Aldo21}.

One of the most used techniques for accomplishing these estimations is the so-called empirical relations among $M$ or $R$ and other stellar parameters, such as: the effective temperature ($T_{\rm eff}$), stellar metallicity ([Fe/H]), (logarithm of) surface gravity ($\log g$), stellar luminosity ($L$), etc. The history of these empirical relations starts in the early twenty century with the works of \citet{Hertzsprung, Russell, Eddington}. Many works have been published since then offering different empirical relations for these estimations or using them for different purposes \citep{Torres10, Gafeira12, Eker15, Benedict16, Eker18, Mann19, Eker21, Fernandes21}. All of them used EB data as the training set for their relations.

Recently, \citet{Moya18} constructed a comprehensive data set including masses and radii obtained from EB and Ast (and only a few interferometric radii). This set consist in a total of more than 700 Main Sequence stars with accurate $M$, $R$, $T_{\rm eff}$, $L$, $\log g$, and [Fe/H]. It was the base of a complete analysis of all the empirical relations possible for estimating $M$ and $R$ as a linear combination of any of the rest of the observables. They presented a total of 38 empirical relations with the best accuracy and precision possible, using classic linear regressions. In addition, they also presented results coming from a Random Forest model as a demonstrator of the benefits of using machine learning techniques for this problem.

This paper is the following step of that work. Taking the advantage of the exceptional database gathered by \citet{Moya18}, we have used state-of-the-art Artificial Intelligence (AI) techniques for getting the most of them to estimate the most accurate stellar masses and radii possible from empirical data. These techniques have the exceptional advantage of extracting the most from the training data, in some cases minimizing the possible biases. We mist take into account that, since the regression models are trained with inputs from EB, Ast, and interferometry (in a few cases), their estimations mimic the results that should be obtained using these techniques. On the other hand, not all of them offer an appropriate treatment and propagation of uncertainties. That is, they use as input information only the observed values and not their uncertainties. In any case, in our opinion, the exercise and model we propose is important and provides valuable information since accurate estimations of stellar masses and radii, regardless of their uncertainties, are an important intermediate step for the statistical analysis of other astrophysical properties, especially in the era where thousands or tens of thousands of observations are analyzed at the same time.

\section{Data sample}
\label{sec:sample}

The data sample is presented and described in detail in \citet{Moya18}. Here we offer a summary of its main characteristics with an impact on our new results. We refer the reader to the original work for additional details about the different sources, data quality, etc.

The data sample consists of 726 Main Sequence stars covering spectral types from B to K with the main bulge of them being F-G stars (an 80 $\%$). All of them have precise estimations of $M$, $R$, $T_{\rm eff}$, $L$, $\log g$, [Fe/H], and stellar mean density ($\rho$, not used in this work). 67 $\%$ of these stars have been characterized using asteroseismology, 31 $\%$ of them belong to EB systems, and the rest (2$\%$) has been studied using interferometry. Asteroseismic stars are mainly solar-like pulsators, explaining the large presence of F-G-type stars in the sample. EB stars are more homogeneously distributed along the HR Diagram (See Fig. 1 in \citet{Moya18} for a detailed distribution of the different subsamples).

In terms of uncertainties, despite the fact that the sample has been gathered from many different sources, masses and radii have been estimated with uncertainties lower than a 7$\%$, and the luminosity has an uncertainty of the order of 10 $\%$. The uncertainties of $T_{\rm eff}$, $\log g$, and [Fe/H] are more heterogeneous but always within the standard values for these observables, that is, around 100 K or less for $T_{\rm eff}$, 0.05 dex or less for $\log g$, and 0.15 dex or less for [Fe/H]. Despite the different HR distribution of the asteroseismic and EB subsets, we have not identified other biases in this data sample.

During the learning step of the AI models we describe in the next section, we take advantage of the error bounds provided in the data sample for each feature. Basically, we proceed to artificially increase the training data with a uniform sampling in the intervals defined by the error bounds, hence generating more samples. For the experiments, we have generated 10 additional samples per original sample. We have observed that generating more samples than that have almost no influence on the results, and slows down the training process. Only neural networks and stacking, those techniques needing a large data sample, would benefit from a larger sample, but not the rest of the models in terms of accuracy.

The inclusion of these additional samples doesn’t change the distribution of the original one. We follow a simple uniform distribution within the error bounds provided in the original data sample to generate additional possible values.  This is a conservative choice because we are assuming that samples can arrive equiprobably over the entire range defined by the uncertainties. We discarded using a Gaussian distribution to generate the samples because it would concentrate most of the samples on the ones currently provided. The presence of not symmetric error bounds makes the employing of a uniform model more appropriate.

\section{Artificial Intelligence Models}
\label{sec:ai_models}
The problem of estimating the mass or radius of stars is primarily a regression problem. Traditionally, the problem has been approached by trying to define empirical relations, which take as input the observables of the stars, to make an estimate of the mass and radius \citep[see][]{Moya18}. We here propose to tackle the problem from a different perspective, replacing this manual search for empirical relations with an approach based on AI models. The idea is simple: considering the observables of the stars as the input features, we let the AI solutions learn the best regression models for estimating $M$ and $R$.

To the best of our knowledge, no similar previous study has been performed with a data sample of this kind. For this reason, in this work, we do not restrict ourselves to a particular AI model but analyze a battery of AI solutions that represent the state of the art. Our goal is to provide a comparative study to elucidate which AI models are most appropriate to address the regression of the mass and radius of stars. In this section, we offer a description of each of the AI regression approaches used in the experiments.

\subsection{Mathematical Notation}
Here we include the mathematical notation used along the description of the AI models proposed. From the data sample described in Section \ref{sec:sample}, we consider that we are given a set of training star feature vectors $\{(\vb{x}_1, t_1), \ldots, (\vb{x}_N, t_N)\}$, such that $\vb{x}_i \in \mathbb{R}^4$ encodes the stellar properties provided in the sample (\ie $T_{\rm eff}$, $L$, $\log g$, [Fe/H]), and $t_i \in\mathbb{R}$ is the corresponding target value we want to estimate. In our case, $t$ can be either the mass ($M$) of the star, or it radius ($R$).

\subsection{Linear regression}

We first propose to learn a Linear regression (LR). Technically, LR assumes that the target ($M$ or $R$) can be estimated employing a linear combination of the features provided in the data sample. Let $\hat{t}$ be the predicted target value, then
\begin{equation}
\hat{t}(\vb{w}, \vb{x}) = w_0 + w_1 x_1 + ... + w_D x_D \; ,
\end{equation}
where vectors $\vb{x}$ and $\vb{w}$ encode the input feature vector (directly taken from the data sample) and the coefficients of the linear regressor, respectively. We follow the least-squares approach \citep{Bishop06} for learning the model, where the objective consists in minimizing the residual sum of squares between the observed targets in the dataset, and the targets predicted by the linear approximation. This is done solving a problem of the form: $\min_{\vb{w}} || \vb{X} \vb{w} - \vb{t}||_2^2$. $\vb{X}$ is a matrix containing all the training vectors, and $\vb{t}$ encodes the target vectors. Similar regression models have been used before, \eg by \citet{Moya18}, thus we include this machine learning technique for validation and comparison purposes.

\subsection{Bayesian Regression}

We implement here a Bayesian regression model to estimate a probabilistic model for the stellar targets. In particular, we follow a Bayesian Ridge Regression (BRR) \citep{Bishop06} of the form:
\begin{equation}
p(t|\vb{X},\vb{w},\alpha) = \mathcal{N}(t|\vb{X} \vb{w},\alpha)\; ,
\end{equation}
where the target $t$ is assumed to be a Gaussian distribution over $\vb{X}\vb{w}$. $\alpha$ is estimated directly from data being treated as a random variable. The regularization parameter is tuned to the data available, introducing over the hyperparameters of the model, \ie $\vb{w}$, the following spherical Gaussian prior $p(\vb{w}|\lambda) =
\mathcal{N}(\vb{w}|0,\lambda^{-1}\mathbf{I}_{D})$.

During model fitting, we jointly estimate parameters $\vb{w}$, $\alpha$ and $\lambda$. Regularization parameters $\alpha$ and $\lambda$ are estimated to maximize the log marginal likelihood following \cite{Tipping01}.

\subsection{Regression Trees and Random Forest} 

We propose to use Decision Trees \citep{Breiman1984} as a non-parametric supervised learning model for the regression of our targets. As they are described by \citet{Breiman1984}, a Regression Tree (RT) estimates a target variable $t$ using a decision tree where a regression model is fitted to each node of the tree to cast the predictions. Different functions can be used to measure the quality of every split performed by the tree nodes (mean squared error, mean absolute error, etc.). 

We also propose to use the AI model known as Random Forest (RF) \citep{Breiman2001}. A Random Forest for regression is a meta estimator that fits various RTs on different sub-samples of the dataset and utilizes averaging to increase predicted accuracy and control over-fitting. Specifically, this AI ensemble model is implemented in our work following the original approach of \citet{Breiman2001}, where the regression trees are built from data samples drawn with replacement from the training set. The mean predicted regression targets of the trees in the forest are used to calculate the predicted regression target of an input sample.

During the learning of RT, we let our model optimizes its internal hyperparameters via a grid search process with cross-validation. For RT, we adjust: the strategy to choose the split at each tree node (best or random); the minimum number of samples required to be at a leaf node (5, 10, 50, 100); and the function to measure the split quality (mean squared error, mean squared error with Friedman’s improvement score for potential splits, mean absolute error or reduction in Poisson deviance to find splits). In the case of RF, we use the following hyperparameters: number of trees (100); minimum number of samples required to be at a leaf node (1); and the function for measuring the quality of the split (mean squared error).

\subsection{Support-Vector Regression}

In machine learning, Support-Vector machines (SVMs) \citep{Boser1992} are one of the most robust supervised learning models. Being originally formulated for classification purposes, they were extended to tackle regression problems by \cite{Drucker1996}, defined as Support-Vector Regression (SVR). We propose to use the SVR AI model for our regression problems. Technically, we follow the $\epsilon$-SVR approach using the LibSVM implementation for regression \citep{LIBSVM}.

Given a set of training vectors, $\{(\vb{x}_1, t_1), \ldots, (\vb{x}_N, t_N)\}$, the goal of $\epsilon$-SVR consists in finding a function that has at most $\epsilon > 0$ deviation from the obtained targets $t_i$ for all our training data, and at the same time is as flat as possible. In other words, we don't mind if regression errors are less than $\epsilon$, but any variation greater than this is unacceptable.

Therefore, the main task in training an SVR is to solve the following optimization problem:

\begin{align}
\label{eq:svr}
 \min_{\vb{w},b,\vb*{\xi},\vb*{\xi^*}} & \Big(\frac{1}{2} \vb{w}^T\vb{w} + C \sum_{i=1}^N\xi_i + C \sum_{i=1}^N \xi_i^*\Big) , \\
 \text{subject to} \; & \vb{w}^T \phi(\vb{x}_i) + b - t_i \leq \epsilon + \xi_i ,  \nonumber\\
 & t_i - \vb{w}^T \phi(\vb{x}_i) - b \leq \epsilon + \xi_i^* ,  \nonumber\\
 & \xi_i, \xi_i^* \geq 0,  i = 1, \ldots, N. \nonumber
\end{align}

$\vb{w}$ defines the hyperplane that best fits the training samples. Flatness for $\vb{w}$ means to seek a small $\vb{w}$, as it is done in Eq. \ref{eq:svr}. 
The function we are searching for is $\vb{w}^T \phi(\vb{x}_i) + b$, being $b$ the bias. 
$\phi(\vb{x}_i)$ is a function that performs a mapping of training vector $\vb{x}_i$ into a higher dimensional space.
$C>0$ is the penalty parameter of the error term, the higher this parameter, the less regression errors we will tolerate.
Finally, $\xi_i$ and $\xi_i^*$ are known as the slack variables. They cope with otherwise infeasible constraints of the optimizaiont problem. That is, thanks to them we allow for some errors (higher than $\epsilon$).

One commonly tackles the following dual problem due to the high dimensionality of the vector variable $w$: 
\begin{equation}
\label{eq:svr_dual}
\min_{\vb*{\alpha}, \vb*{\alpha}^*} \;\Big[ \frac{1}{2} (\vb*{\alpha} - \vb*{\alpha}^*)^T Q (\vb*{\alpha} - \vb*{\alpha}^*) + \epsilon \sum_{i=1}^N (\alpha_i + \alpha_i^*) + \sum_{i=1}^N t_i (\alpha_i - \alpha^*_i) \Big] ,
\end{equation}
\vspace{-0.75cm}
\begin{align}
\text{subject to} \; & \vb{e}^T (\vb*{\alpha} - \vb*{\alpha}^*)= 0, \nonumber\\
& 0 \leq \alpha_i, \alpha_i^* \leq C, i = 1, \ldots,N, \nonumber
\end{align}
where $\vb{e}=[1, \ldots, 1]^T$ is a vector of all ones, $Q$ is an $N \times N$ positive semi-definite matrix, and $Q_{ij} = K(\vb{x}_i,\vb{x}_j) \equiv \phi(\vb{x}_i)^T \phi(\vb{x}_j)$. $K(\vb{x}_i,\vb{x}_j)$ is known as the kernel function. For this study we employ the Radial Basis Function (RBF) kernel because it is the one reporting the best results. The RBF kernel is defined as follows:
\begin{equation}
K(\vb{x}_i,\vb{x}_j) = \text{exp}{\left( -\frac{||\vb{x}_i - \vb{x}_j ||^2}{2\sigma^2} \right)} \; ,
\end{equation}
where $\sigma$ is a free parameter of the kernel. For our experiments, $\epsilon$ is fixed to 0.1 and we perform a grid search with cross-validation to adjust the regularization parameter $C$ (1, 10, 100, 1000) and $\gamma = \frac{1}{2\sigma^2}$ ($10^{-3}, 10^{-4}$). {\bf The best values found were the following: i) for the estimation of the stellar mass, $C=10$ and $\gamma=0.001$; ii) for the radius estimation, $C=1000$ and $\gamma=0.001$.}

\subsection{kNN} 

As an interesting baseline, we also include in this study the k-Nearest Neighbor (kNN) regression model. The output of a kNN regression model is obtained as the average of values of the targets of the k nearest neighbors for the test input. Technically, given a test sample $\vb{x}_i$, our model first identifies the k nearest neighbors in the training set. This is done employing the Euclidean distance. Then, the estimation for the test $\vb{x}_i$ is obtained as the the mean of the targets of its k nearest neighbors. For our experiments, we test the following k parameter values: 1, 5, 10, 15, 20, and 50, being k=5 that reporting the best results on average for all the experiments and targets. Therefore, we fix parameter k to 5. The type of weight function used to scale the contribution of each neighbor is uniform, which means that all points in each neighborhood are weighted equally.

\subsection{Neural Networks}

We analyze in this study the performance of Neural Networks (NNs) for our problem. Due to the size of the database we handle, and the type of input data, we have decided to use deep feedforward networks models \citep{Goodfellow2016}. Technically, we have implemented a multi-layer perceptron to learn a mapping of the form $t=f(\vb{x},\vb{w})$, where $\vb{w}$ encodes model parameters. Our NN is able to learn the values for $\vb{w}$ that result in the best function approximation in the form of a feedforward network.

\begin{figure}[t]
\begin{center}
\includegraphics[width=0.6\linewidth]{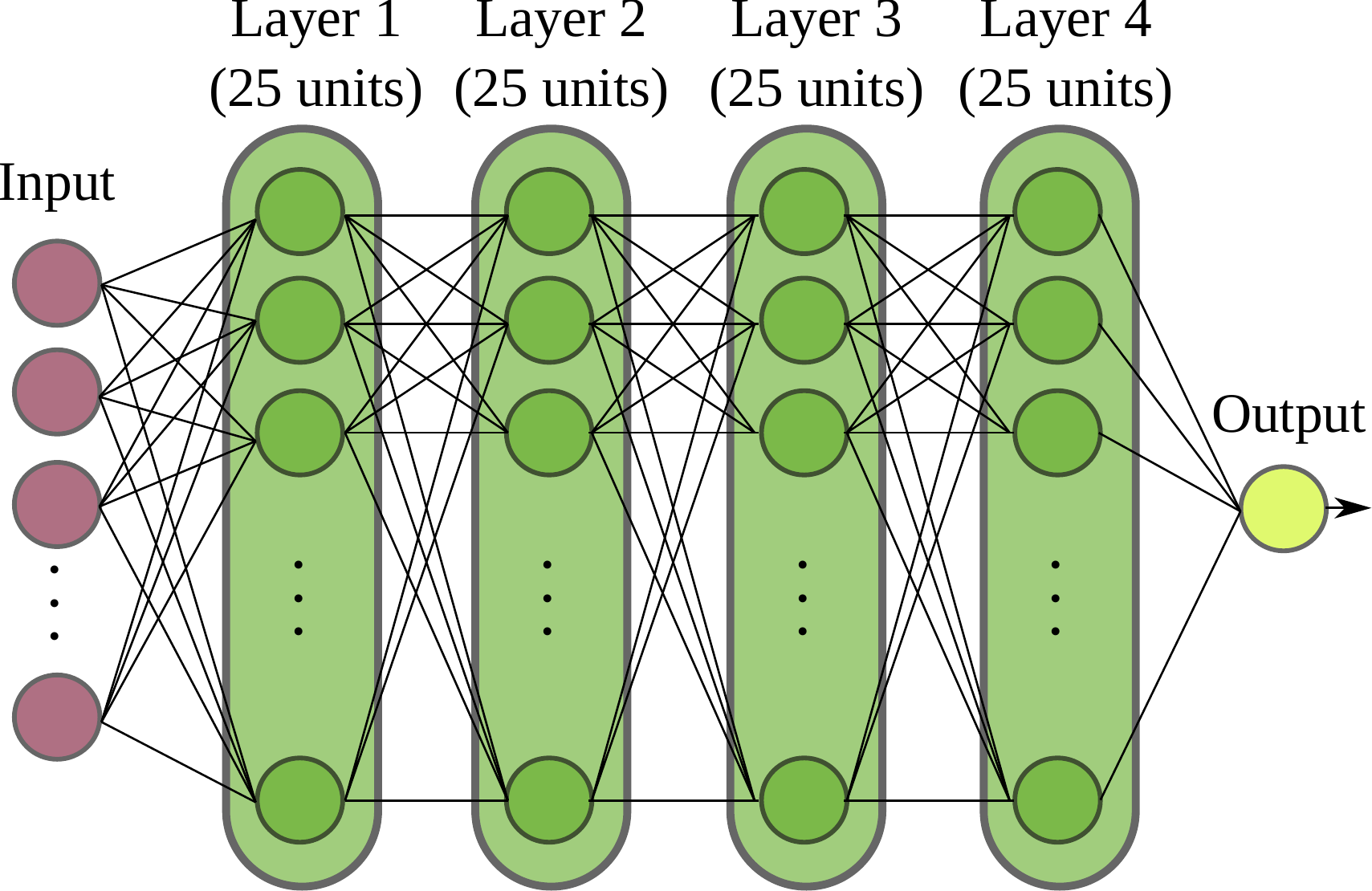}
\end{center}
\caption{Architecture of the feedforward neural network implemented. 4 hidden layers have been used with 25 units each of them followed by a ReLU. The output layer has no activation function to directly perform the star parameter regression.}
\label{fig:neural_network}
\end{figure}

The architecture implemented in our work is depicted in Figure \ref{fig:neural_network}. It consists of a multi-layer perceptron with an input layer, a set of 4 hidden fully-connected layers with 25 units each, and the output layer in charge of the regression of the target variable. Every hidden unit is followed by a ReLU activation function. We use as loss function the square error, 
\begin{equation}
Loss(\hat{t},t,\vb{w}) = \frac{1}{2}||\hat{t} - t ||_2^2 + \frac{\alpha}{2} ||\vb{w}||_2^2 \; ,
\end{equation}
where $\hat{t}$ is the prediction of the NN, and $\alpha$ is the regularization parameter. Backpropagation \citep{LeCun2012} is used for learning the model with SGD~\citep{sgd} optimizer. During learning, we fix $\alpha=0.01$ and use an adaptive learning rate policy. For the estimation of $R$ and $M$, the initial learning rate is fixed to 0.09 and 0.2, respectively, because they provide the best results.

\subsection{Stacking}
\label{sec:stacking}

\begin{figure}[ht]
\centering
\includegraphics[width=0.8\linewidth]{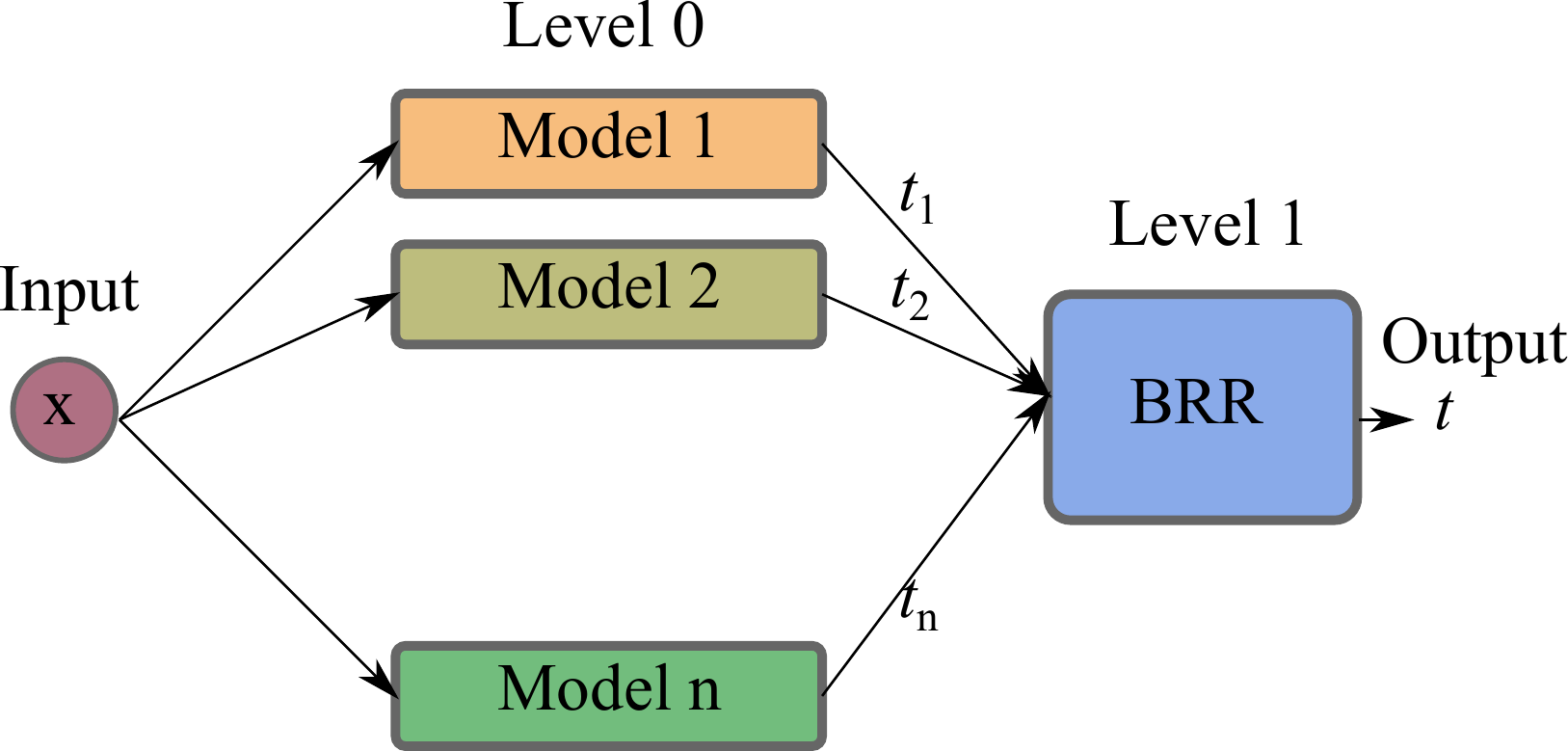}
\caption{Architecture of the stacked generalization model implemented. We propose a specific architecture for each stellar property, where at level 0 we integrate its best regression models. At level 1 both models use a BRR, which is trained on out-samples (taken from the training set), following a cross-validation methodology.}
\label{fig:stacking}
\end{figure}

Finally, we use a machine learning ensemble method known as a stacked generalization (or stacking) \citep{Wolpert1992}. Stacked generalization involves stacking the output of a set of individual estimators (level 0) to learn a final estimator (level 1), see Figure \ref{fig:stacking}. In a regression problem like ours, the predictions of each individual level 0 regression model, \ie $(t_1, t_2,...,t_n)$ in Figure \ref{fig:stacking}, are stacked and fed into a final level 1 estimator, which calculates the final prediction for our targets. Overall, stacking is an ensemble strategy in which the model at level 1 learns how to combine the predictions from multiple existing models in the most effective way possible.
Note that this strategy is different from other ensemble methods like bagging or boosting. Unlike bagging, the models in stacking are usually distinct (\eg, not all decision trees as in a random forest) and fit on the same dataset (\eg. instead of samples of the training dataset).
Different from boosting, stacking uses a single model at level 1 to learn how to integrate the predictions from the contributing level 0 models in the most effective way (e.g. instead of a sequence of models that correct the predictions of prior models). Having said this, the use of stacking allows us to explore and combine a variety of heterogeneous regression models.

We propose a specific stacking architecture for each stellar target variable, where at level 0 we simply integrate its corresponding best regression models. At level 1 both architectures uses a linear BRR. 
Typically, the stacking meta-model at level 1 should be simple, allowing for a smooth interpretation of the base models' predictions. This way the final prediction works as a weighted average or blending of the predictions given by the base models. Using for level 1 more complex models (\eg a NN) yields worse results.
Moreover, we are aware that a proper uncertainty propagation is one of the main requirements for any astrophysical study, and not all the proposed AI techniques are able to naturally provide that. Using a BRR at level 1 allows us to provide a model that can associate to each regression its corresponding uncertainty. Therefore, we have decided to build this approach thinking on a consistent error propagation pipeline, and for that, a bayesian regression is, in our opinion, the best choice for stacking.

During training, and to avoid over-fitting, this final BRR estimator of level 1 is trained on out-of-samples. That is, data \emph{not} used for training the models in level 0 is fed to these $n$ models. Then, predictions $(t_1, t_2,...,t_n)$ at level 0 are made, and used along with the expected target values as training pairs to fit the level 1 model. To prepare the training data for the level 1 model, we follow a standard 5-fold cross-validation \citep{Hastie2009} of the models at level 0, where the out-of-fold predictions are used to train the level 1 model.

\subsection{Evaluation metrics}

For evaluating the performance of the different AI regression models, we propose the following three metrics. For all our experiments, the main evaluation metric used is the Mean Absolute Error (MAE),
\begin{equation}
\rm MAE = \frac{\sum_{i=1}^{N}|t_i-\hat{t}_i|}{N} \; ,
\end{equation}
where $t_i$ and $\hat{t}_i$ are the target (mass or radius) provided in our dataset, and the corresponding value estimated by a regression model, respectively. When we compare the estimated mass (or radius) $\hat{t}_i$ using our AI models with the \emph{real} mass (or radius) $t_i$ for the testing sample, it is also interesting to measure the dispersion of the estimations with respect to the real values. For a numerical estimation of this dispersion, we have calculated the Mean Relative Difference (MRD) and Mean Absolute Relative Difference (MARD), defined as follows,

\begin{equation}
\rm MARD = \frac{\sum_{i=1}^{N}\frac{|t_i-\hat{t}_i|}{t_i}}{N}  \times 100 \; ,
\end{equation}

\begin{equation}
\rm MRD = \frac{\sum_{i=1}^{N}\frac{t_i-\hat{t}_i}{t_i}}{N}  \times 100 \; ,
\end{equation}
where, MRD measures the bias of our estimations, while MARD focuses on their accuracy.

Note that during our study we assume that the dataset provides real masses and radii for the stars in the sample. These features have been determined by any of the techniques described in Section \ref{sec:sample}: \eg asteroseismology, EBs, and interferometry in a few cases. In our study we therefore try to mimic these techniques with machine learning models fed with training data.

\section{Results and discussion}

The problem of stellar mass and radius estimation is considered here as a regression problem. With the following experiments we propose to analyze the suitability of the set of AI models described in the previous section for this particular problem.

In order to reproduce all the results of our study, we release all codes and data used, in the following repository: \url{https://github.com/gramuah/ai4mr}. In addition to the dataset used in the experiments, we publicly release the different AI models that have been employed in this study. We have built all the regression models in Python, using the free software machine learning library scikit-learn (version 1.0) \citep{scikit-learn}, which requires a Python version $\geq 3.7$. Our AI models are also available at the link \url{http://sdc.cab.inta-csic.es/empiricalRelationsMR} for the online estimation of stellar masses and radii. In this last service, the linear regressions of \citet{Moya18} are also provided.

\subsection{AI for stellar mass and radius estimation}
\label{sec:experimental_setup}

\begin{figure}
\begin{center}
\includegraphics[width=\linewidth]{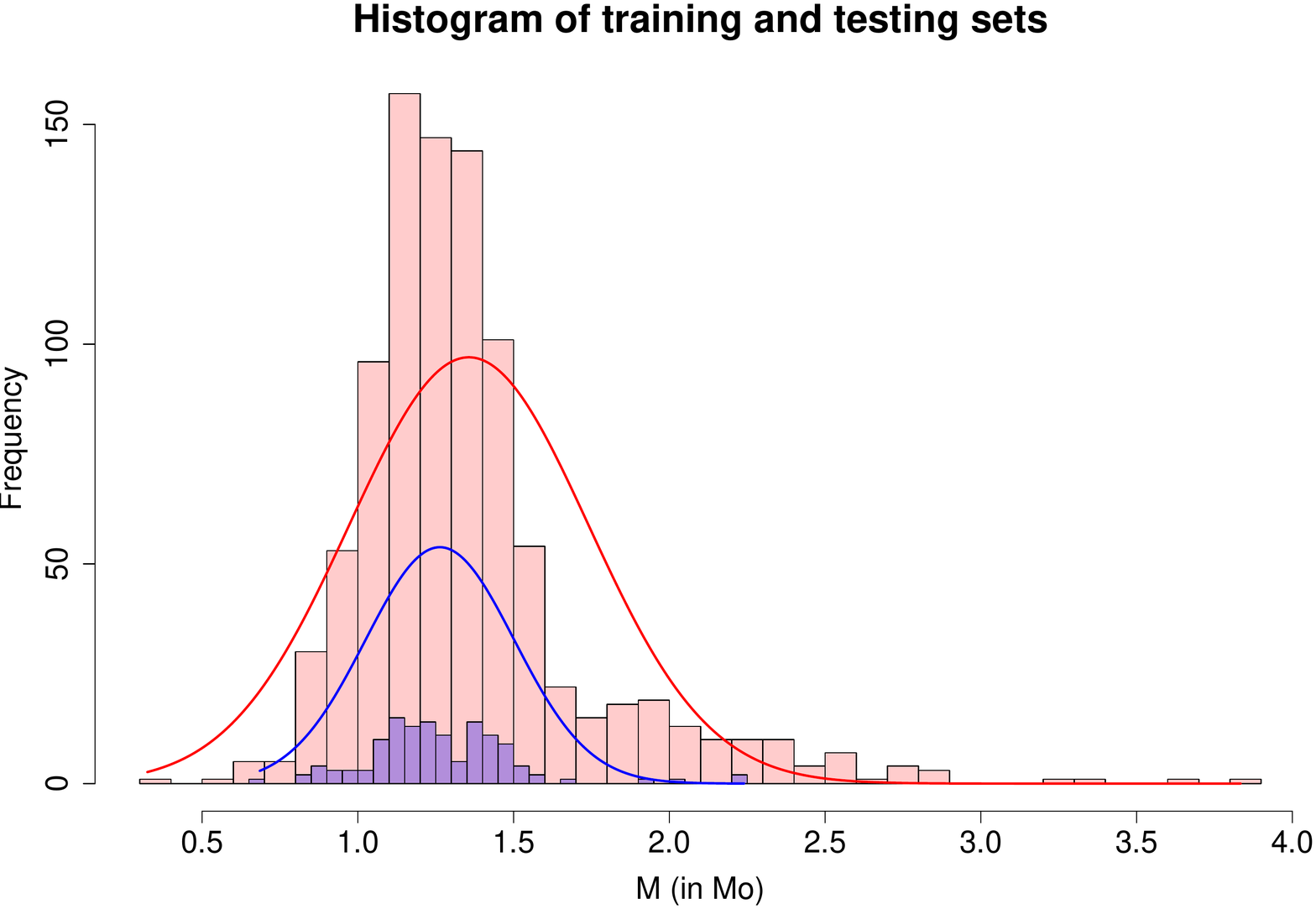}
\caption{Histogram for the masses of the stars included in the training (in red) and the testing (in blue) splits.}\label{fig:hist_M}
\end{center}
\end{figure}

\begin{figure}
\begin{center}
\includegraphics[width=\linewidth]{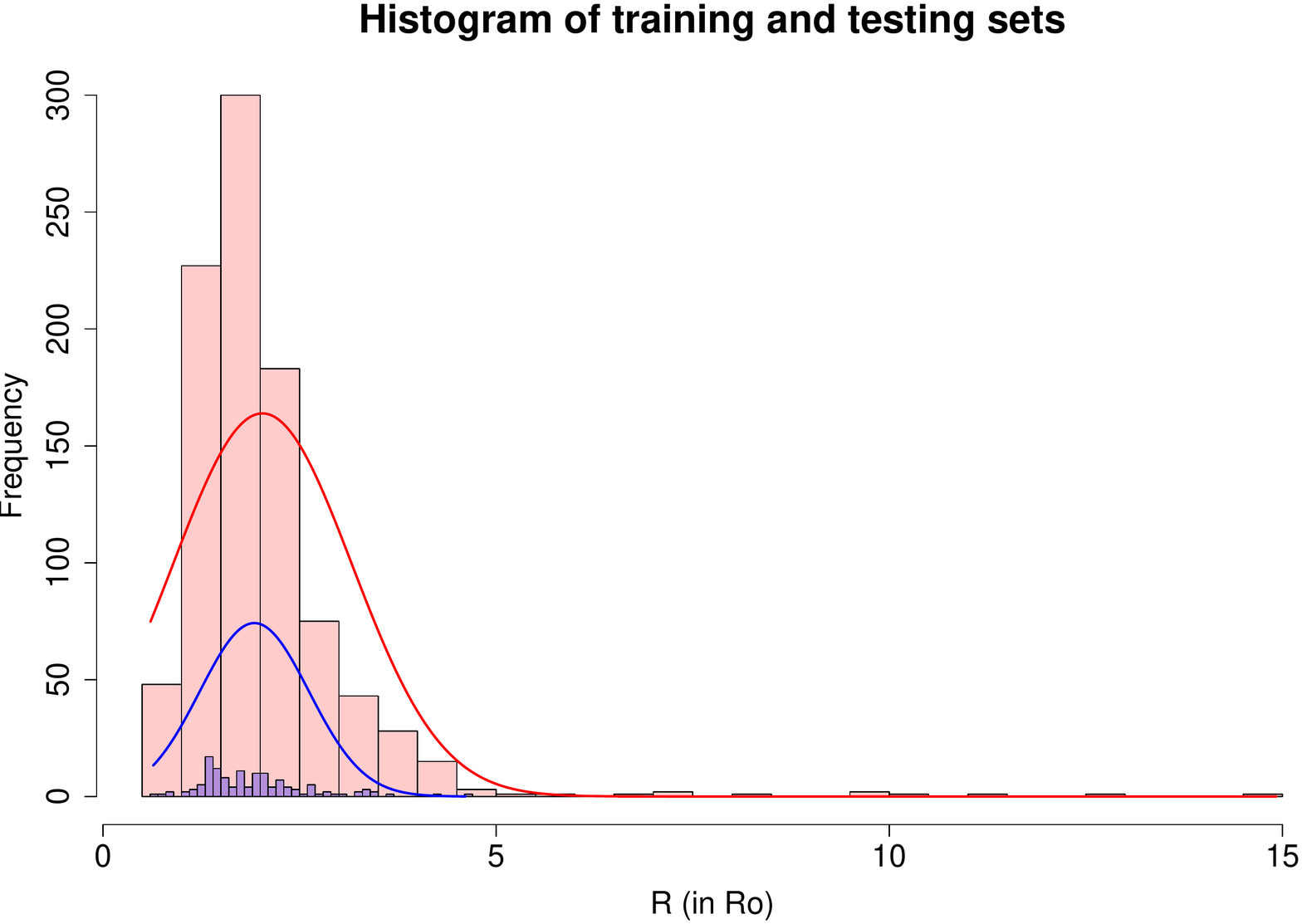}
\caption{Histogram for the radii of the stars included in the training (in red) and the testing (in blue) splits.}\label{fig:hist_R}
\end{center}
\end{figure}

Using the data sample of \cite{Moya18}, we split it into a training and a testing set, the same for all the experiments. We randomly select $80\%$ of the samples for the training set, and the rest of stars, \ie $20\%$ of the sample, are used for testing purposes. Figures \ref{fig:hist_M} and \ref{fig:hist_R} show the histograms for masses and radii, respectively, of the training and testing sets. We have artificially incremented the density curve fitting the testing set for an easier visual comparison with the training set. They show the statistical coherence of the split, since the testing set covers properly the training set population. Our first experiment analyzes the effectiveness of different AI techniques in predicting these test values using the training set to learn them.

In Table \ref{table:ai_models} we show the MAE, on the test set, for all the AI models detailed in Section \ref{sec:ai_models}. If no stacked generalization is used, the best models for both target variables are the neural networks designed. They report an MAE of 0.05 and 0.049 for the mass and radius, respectively. The stacking architectures, which combine for every target variables their best models at level 0, report the lowest errors. In particular, the stacking architecture for the mass integrates at level 0 the following models: NN, kNN and RF. Note that we do not include the SVR at level 0, although it reports a better MAE than the kNN. In our tests, including the SVR did not yield better results, and it also slowed down the training. For the radius, our stacking architecture simply integrates at level 0 the NN and the SVR. We offer a detailed comparison of our two stacking models with the state-of-the-art in Section \ref{sec:sota_comparison}, but we advance here that they have improved the performance of classical techniques based on empirical relations. 

\begin{table*}
\centering
%\scalebox{0.70}{
\caption{MAE for every AI model of section \ref{sec:ai_models} when estimating stellar masses and radii for the testing sample. }\label{table:ai_models}
\begin{tabular}{l|cccccccc}  
\toprule
\textbf{Models} & LR & BRR & RT & RF & SVR & kNN & NN & Stacking \\
\midrule
Mass MAE (in M$_\odot$) & 0.075 & 0.075 & 0.069 & 0.062 & 0.063 & 0.065 & 0.050 & \textbf{0.049}\\
Radius MAE (in R$_\odot$) & 0.128 & 0.128 & 0.076 & 0.069 & 0.050 & 0.093 & 0.049 & \textbf{0.048}\\
\bottomrule
\end{tabular}
%}%end of scalebox
\end{table*}

\begin{figure*}
    \centering
    \begin{subfigure}{0.2\linewidth}
        \includegraphics[width=\linewidth]{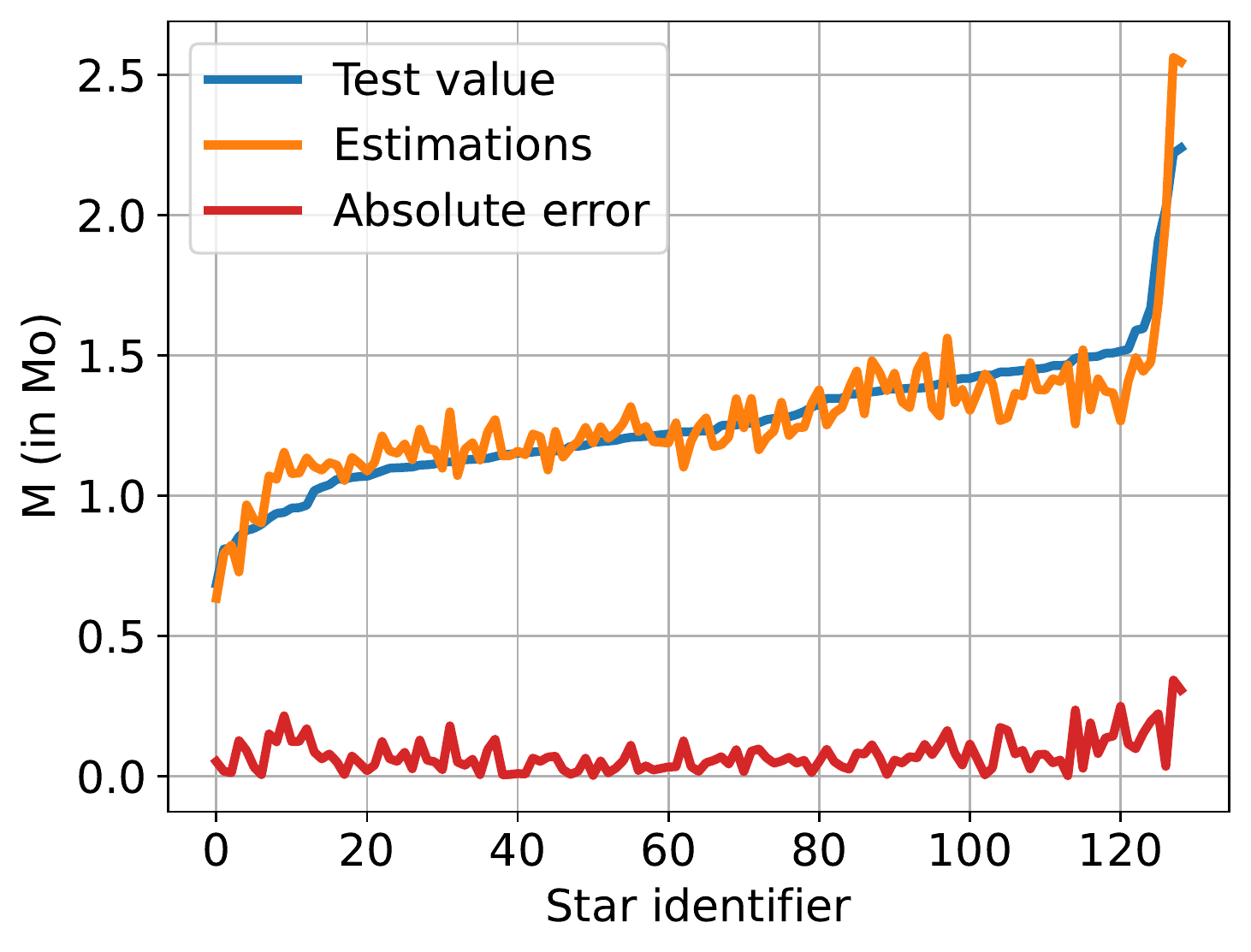}
        \caption{LR}
        \label{fig:ai_models_for_M:lr}
    \end{subfigure}
    \begin{subfigure}{0.2\linewidth}
        \includegraphics[width=\linewidth]{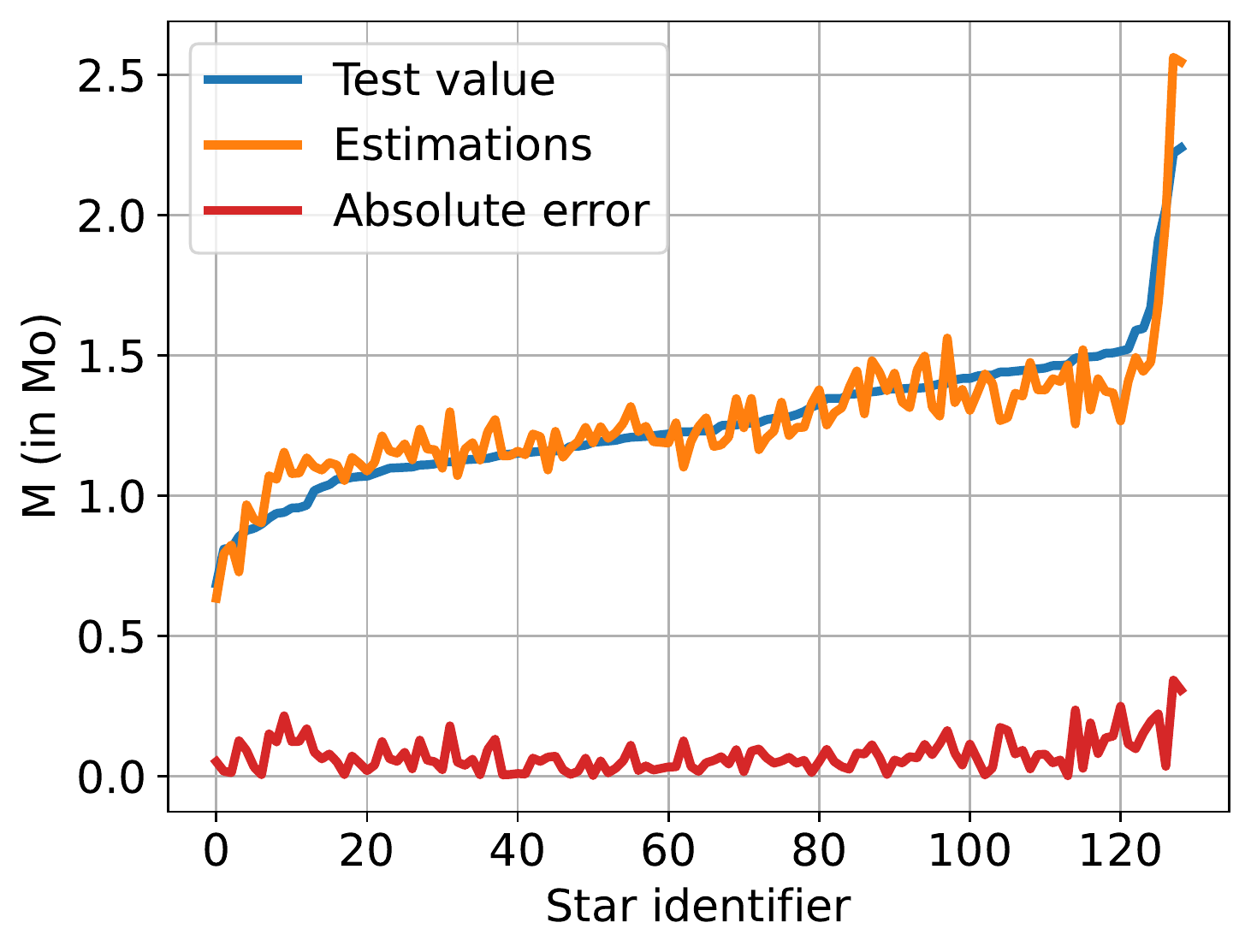}
        \caption{BRR}
        \label{fig:ai_models_for_M:bayes}
    \end{subfigure}
    \begin{subfigure}{0.2\linewidth}
        \includegraphics[width=\linewidth]{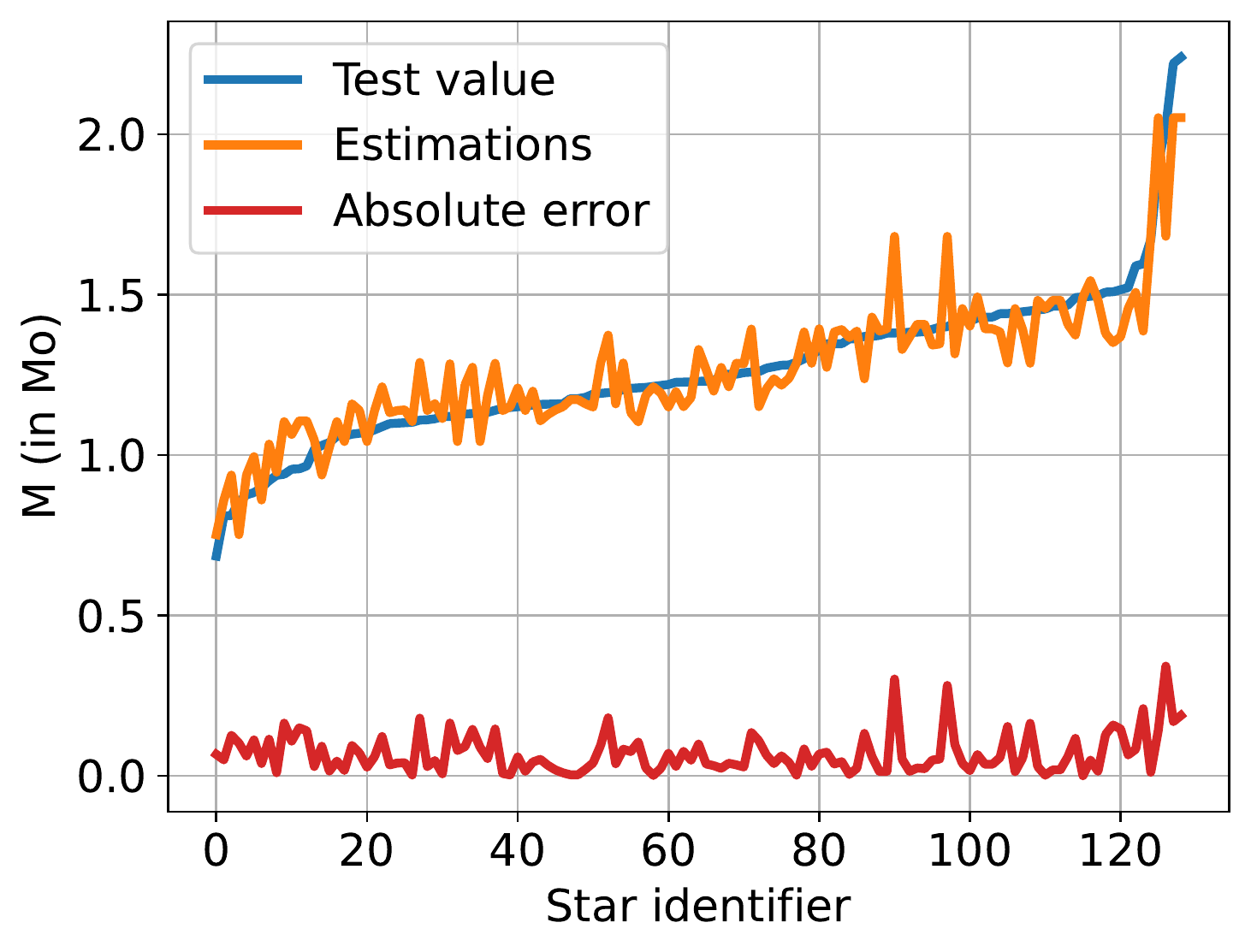}
        \caption{RT}
        \label{fig:ai_models_for_M:dtr}
    \end{subfigure}
    \begin{subfigure}{0.2\linewidth}
        \includegraphics[width=\linewidth]{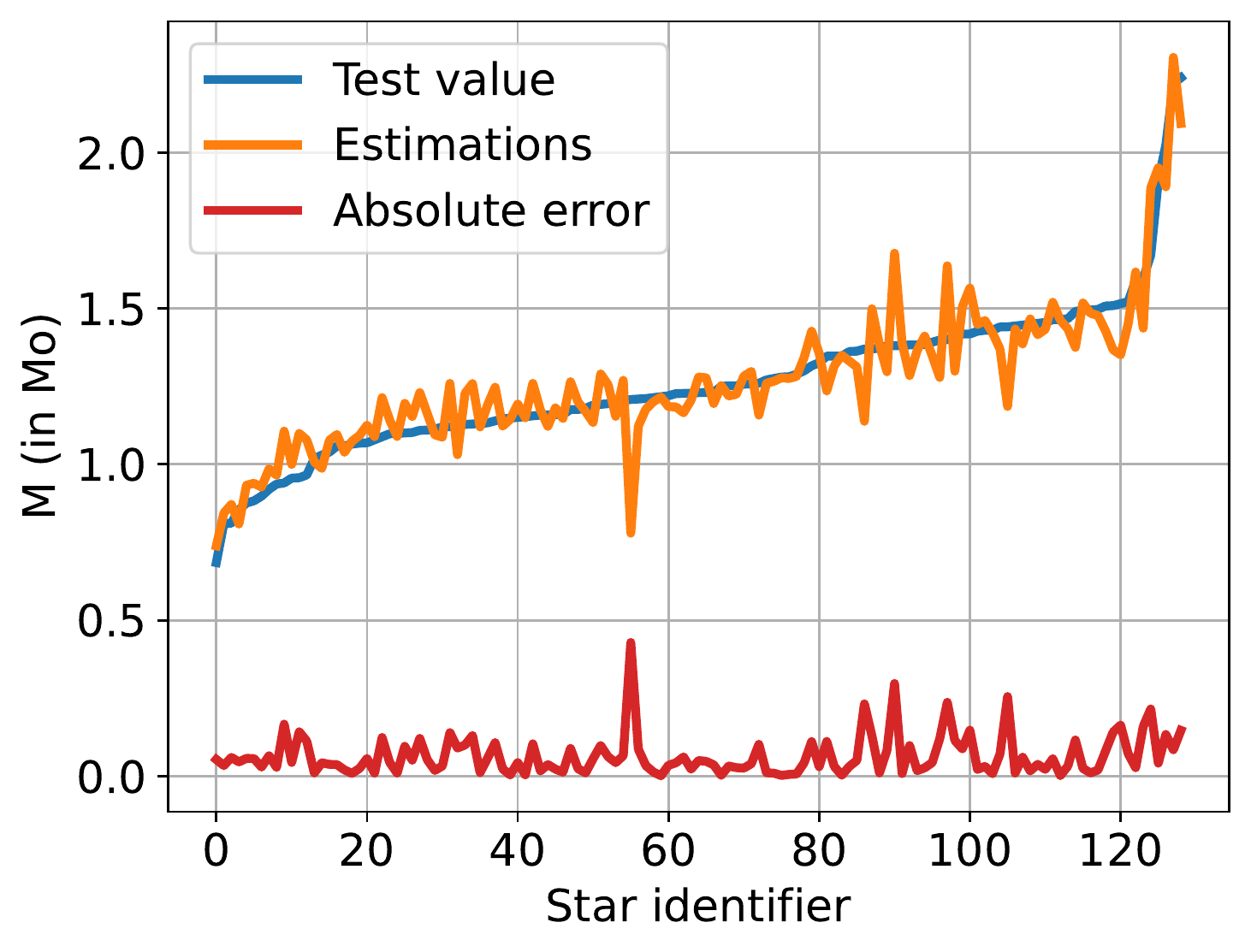}
        \caption{RF}
        \label{fig:ai_models_for_M:rf}
    \end{subfigure}
    %%%%%%%%%%%%%%%%%%%%%%%%%%%%%%%%%%%%%%%%%%%%%%%%%%%%%%%
    \hfill %%%%%%%%%%%%%%%%%%%%%%%%%%%%%%%%%%%%%%%%%%%%%%%%
    %%%%%%%%%%%%%%%%%%%%%%%%%%%%%%%%%%%%%%%%%%%%%%%%%%%%%%%
    \begin{subfigure}{0.2\linewidth}
        \includegraphics[width=\linewidth]{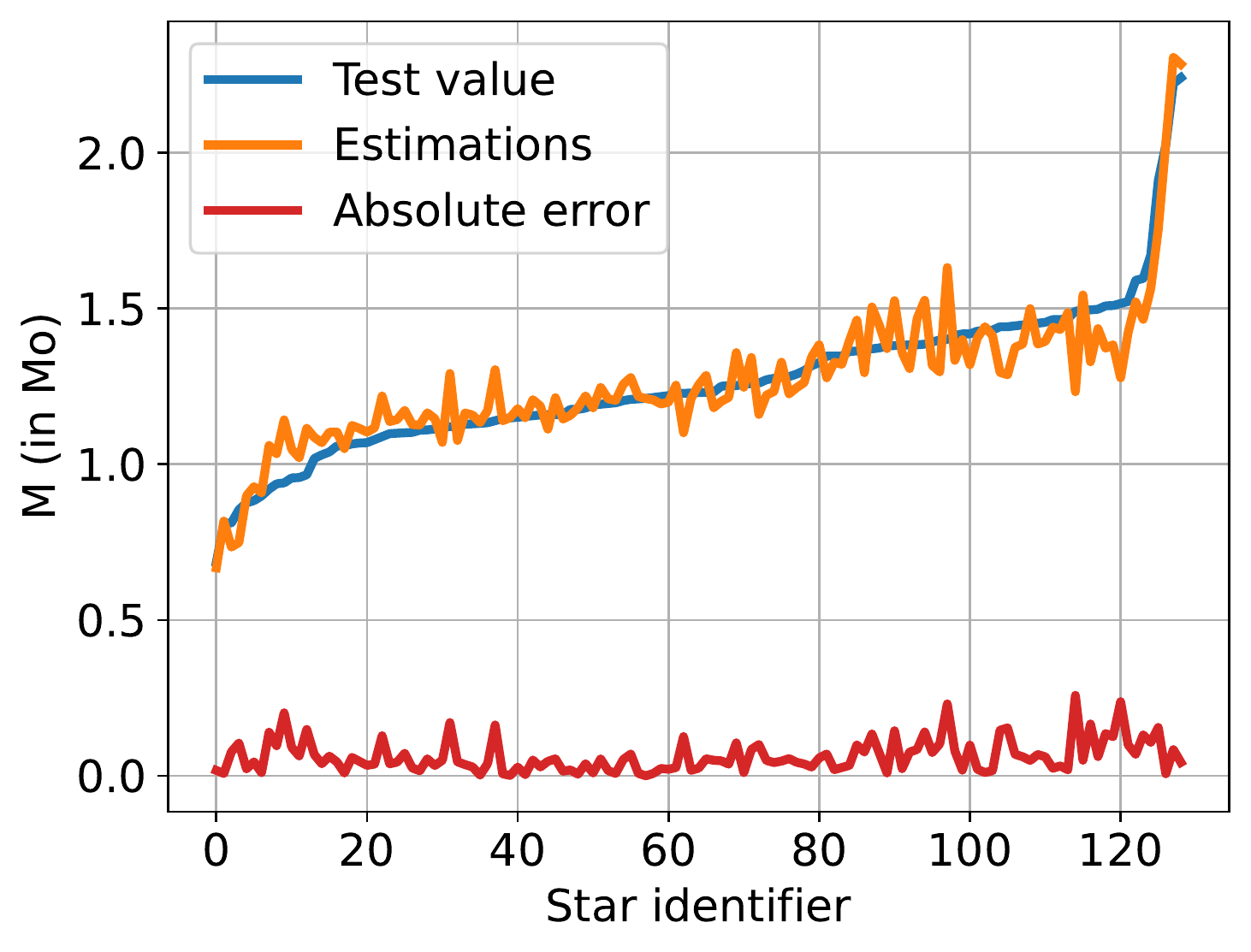}
        \caption{SVR}
        \label{fig:ai_models_for_M:svr}
    \end{subfigure}
    \begin{subfigure}{0.2\linewidth}
        \includegraphics[width=\linewidth]{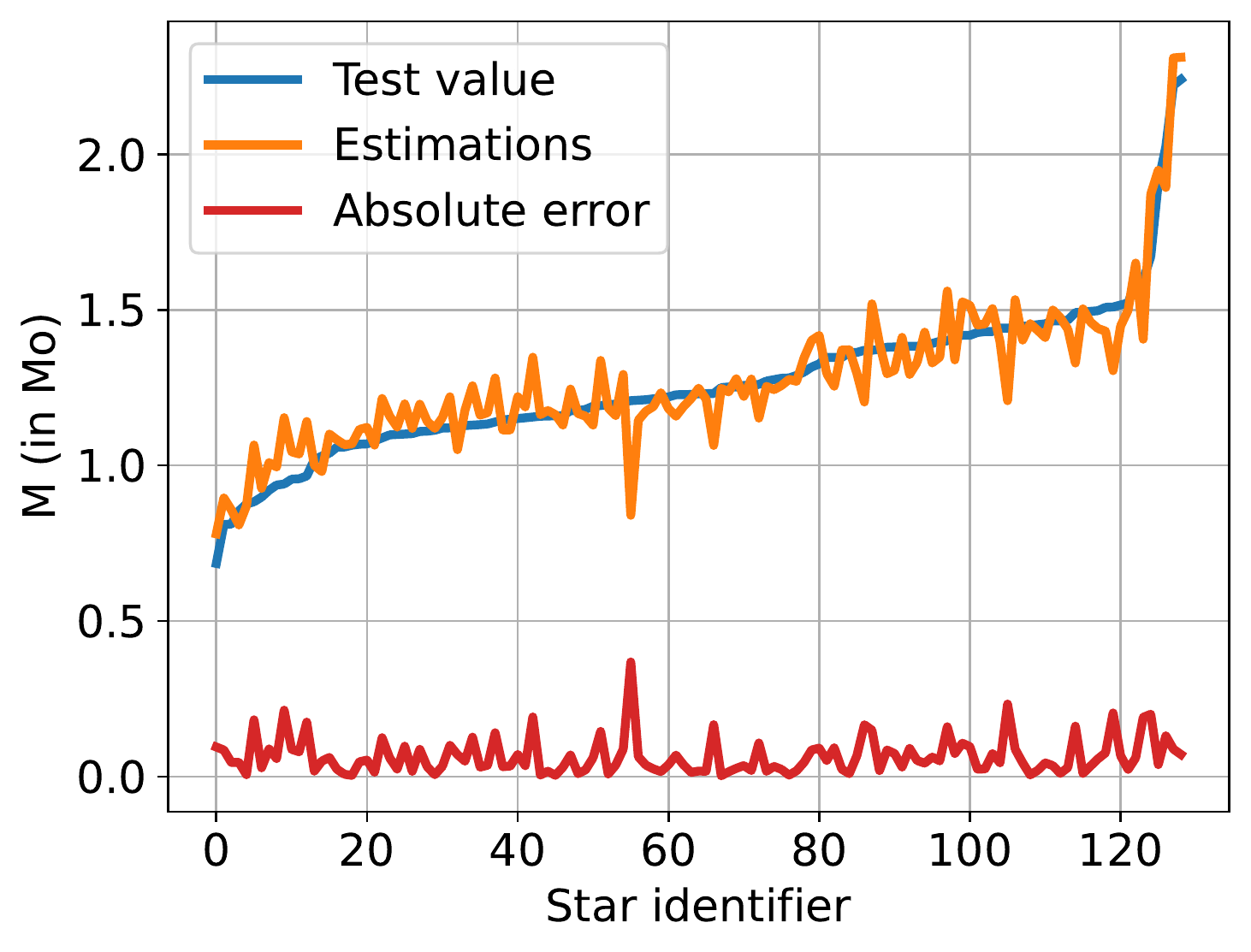}
        \caption{kNN}
        \label{fig:ai_models_for_M:knn}
    \end{subfigure}
    \begin{subfigure}{0.2\linewidth}
        \includegraphics[width=\linewidth]{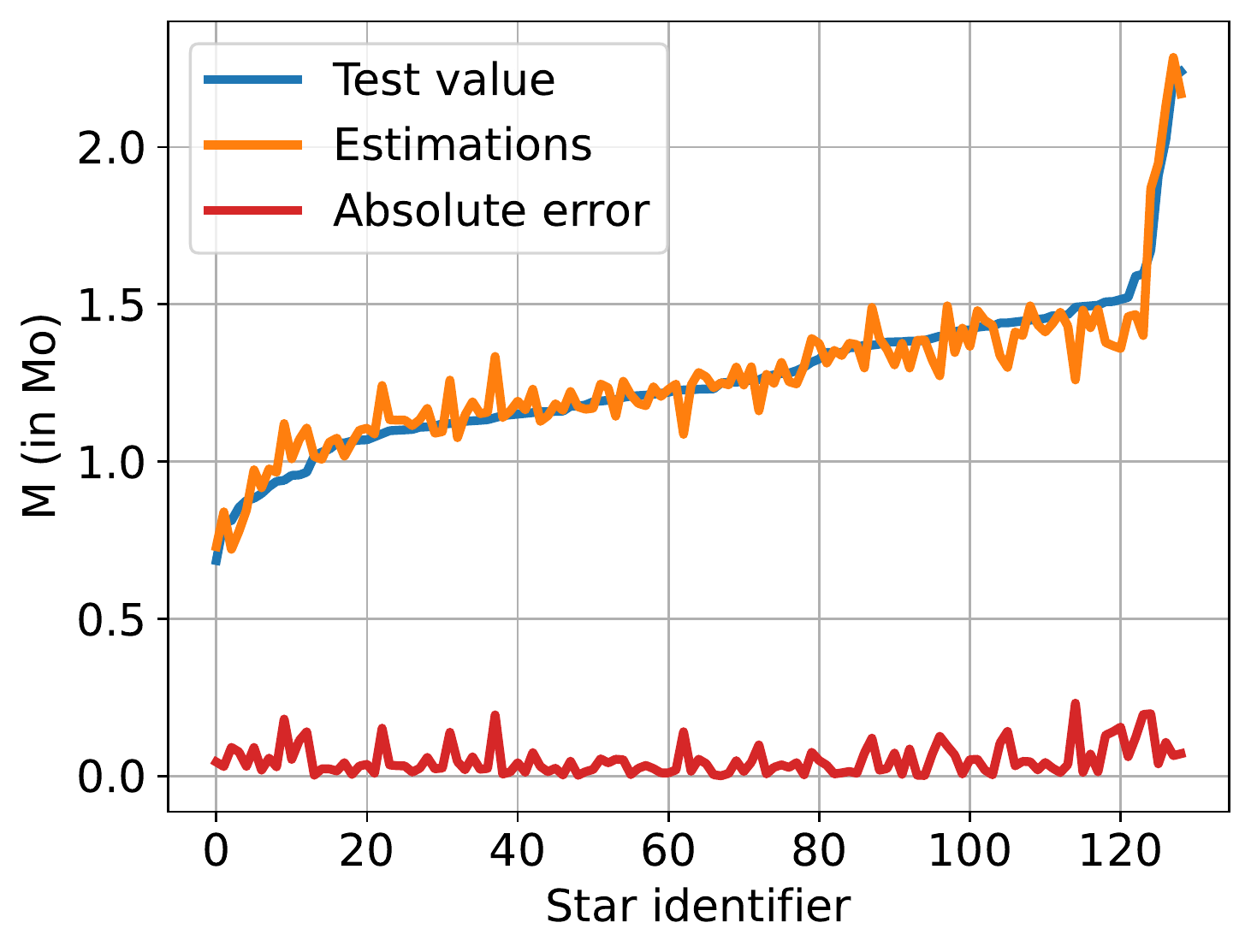}
        \caption{NN}
        \label{fig:ai_models_for_M:nn}
    \end{subfigure}
    \begin{subfigure}{0.2\linewidth}
        \includegraphics[width=\linewidth]{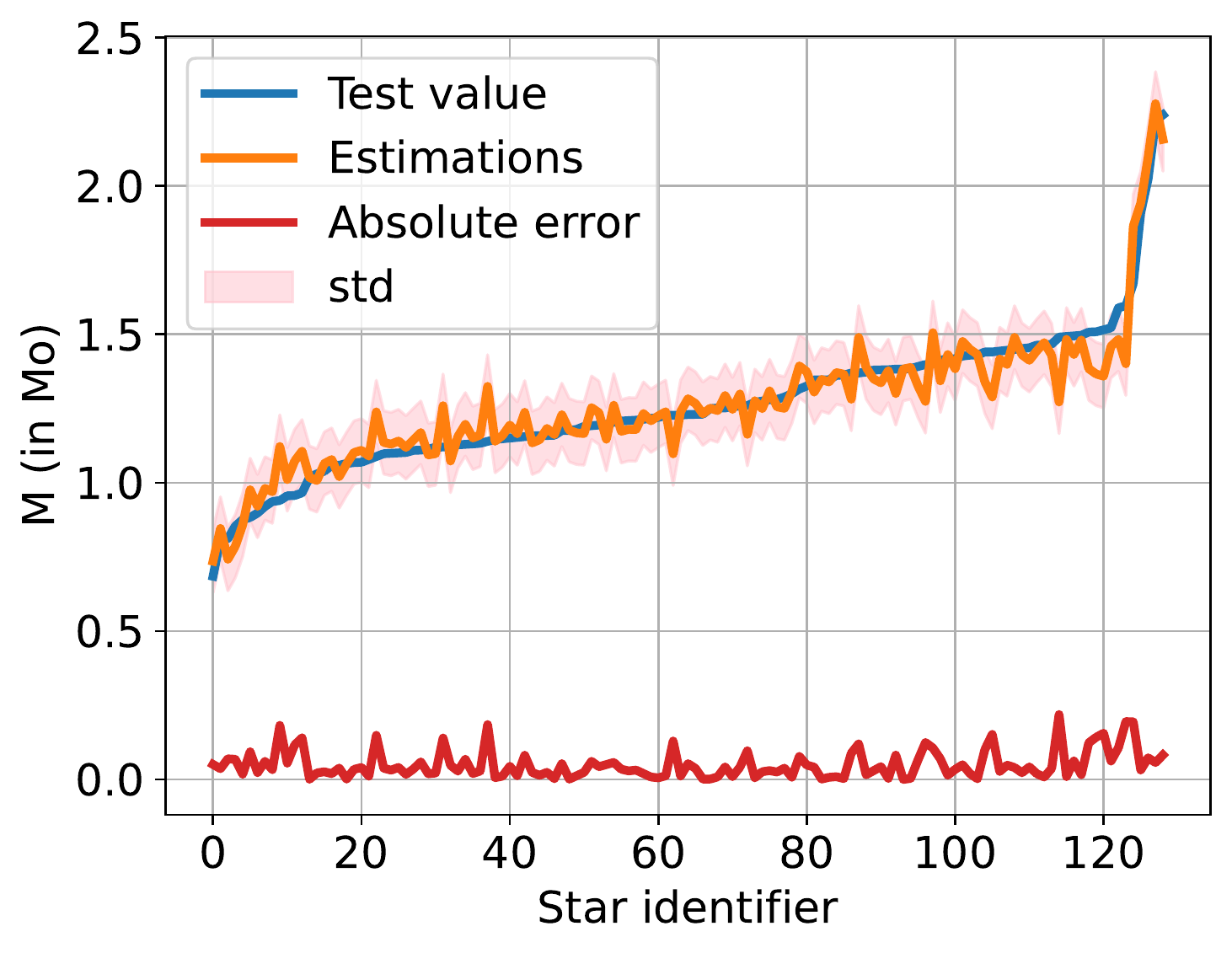}
        \caption{Stacking}
        \label{fig:ai_models_for_M:stacking}
    \end{subfigure}
    \caption{Detailed estimations of the stellar mass for every test sample of all the AI models proposed in this paper. The orange line shows the masses estimated by the AI techniques, and the blue line shows the corresponding test value. The red line represents the absolute error between these two previous quantities. The X-axis shows the identifier of each test star when they have been sorted in increasing order of mass.}
    \label{fig:ai_models_for_M}
\end{figure*}

\begin{figure*}
    \centering
    \begin{subfigure}{0.2\linewidth}
        \includegraphics[width=\linewidth]{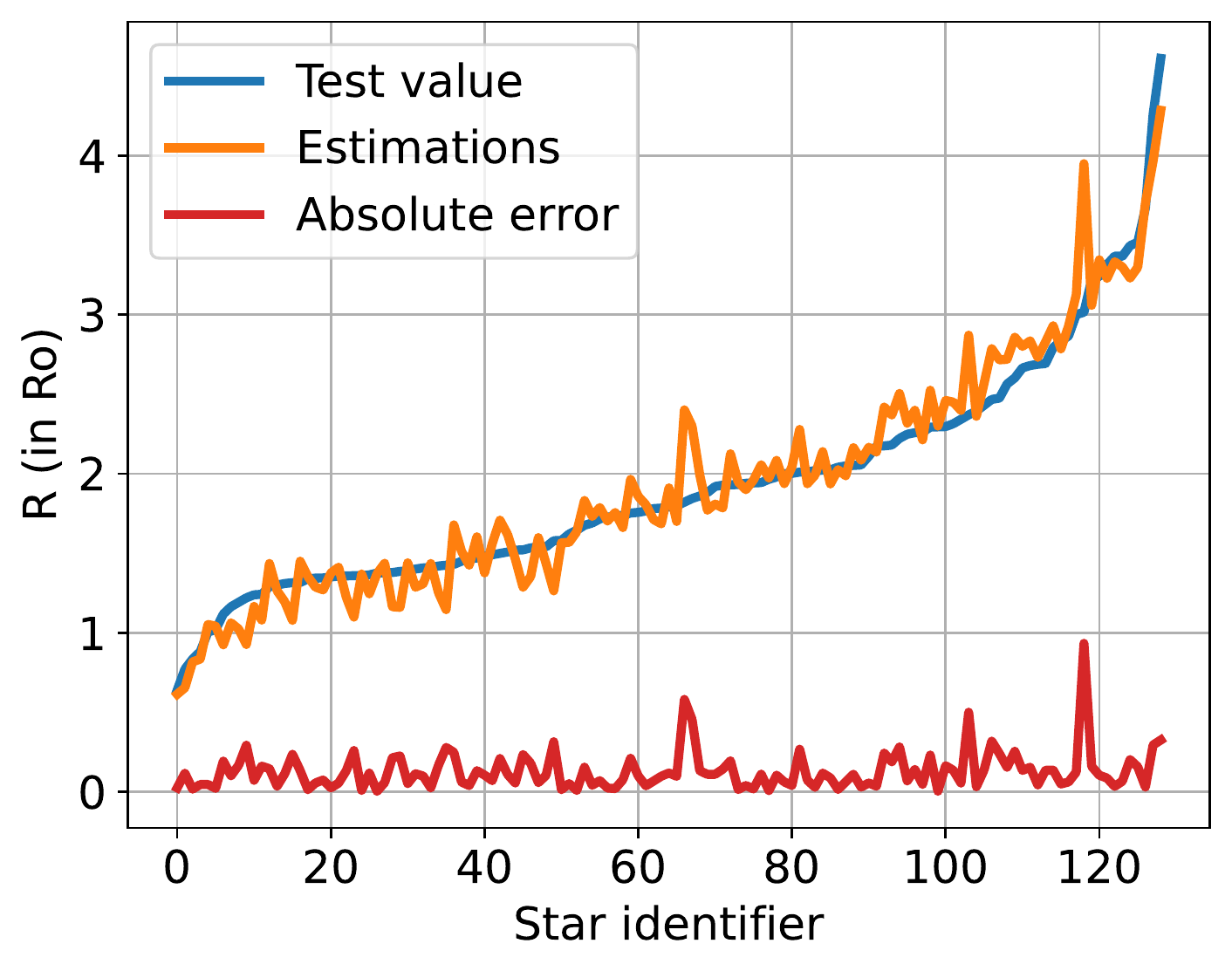}
        \caption{LR}
        \label{fig:ai_models_for_R:lr}
    \end{subfigure}
    \begin{subfigure}{0.2\linewidth}
        \includegraphics[width=\linewidth]{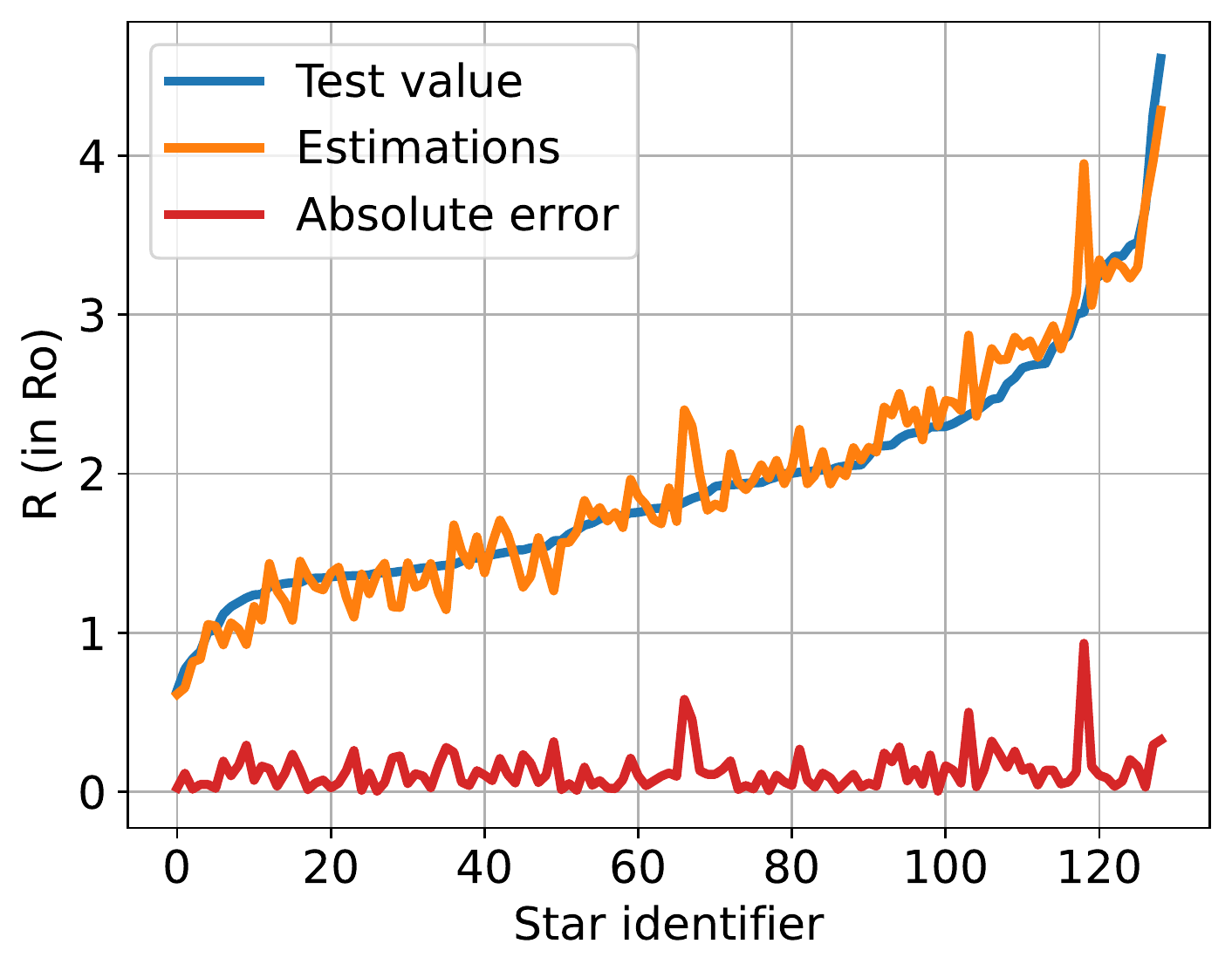}
        \caption{BRR}
        \label{fig:ai_models_for_R:bayes}
    \end{subfigure}
    \begin{subfigure}{0.2\linewidth}
        \includegraphics[width=\linewidth]{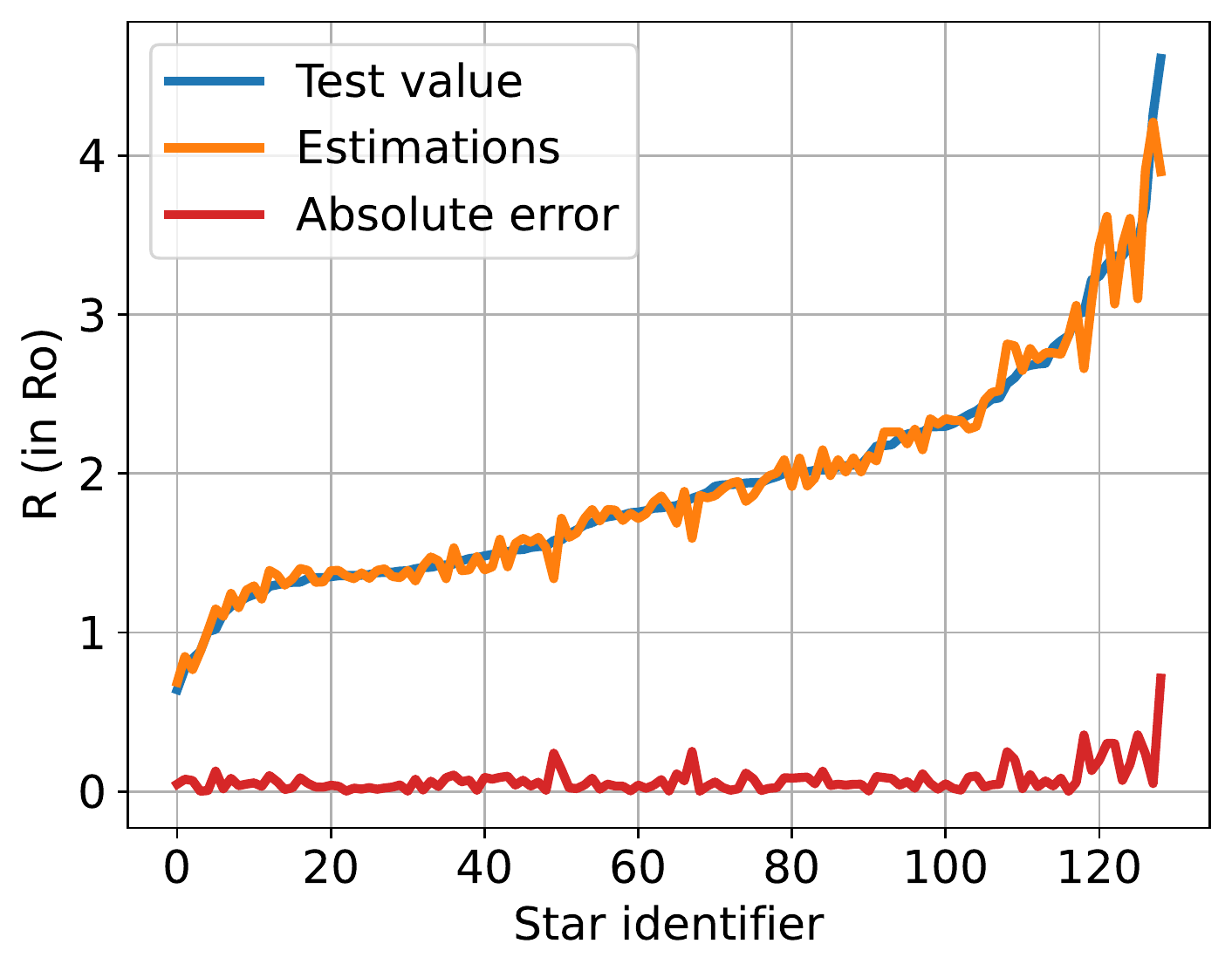}
        \caption{RT}
        \label{fig:ai_models_for_R:dtr}
    \end{subfigure}
    \begin{subfigure}{0.2\linewidth}
        \includegraphics[width=\linewidth]{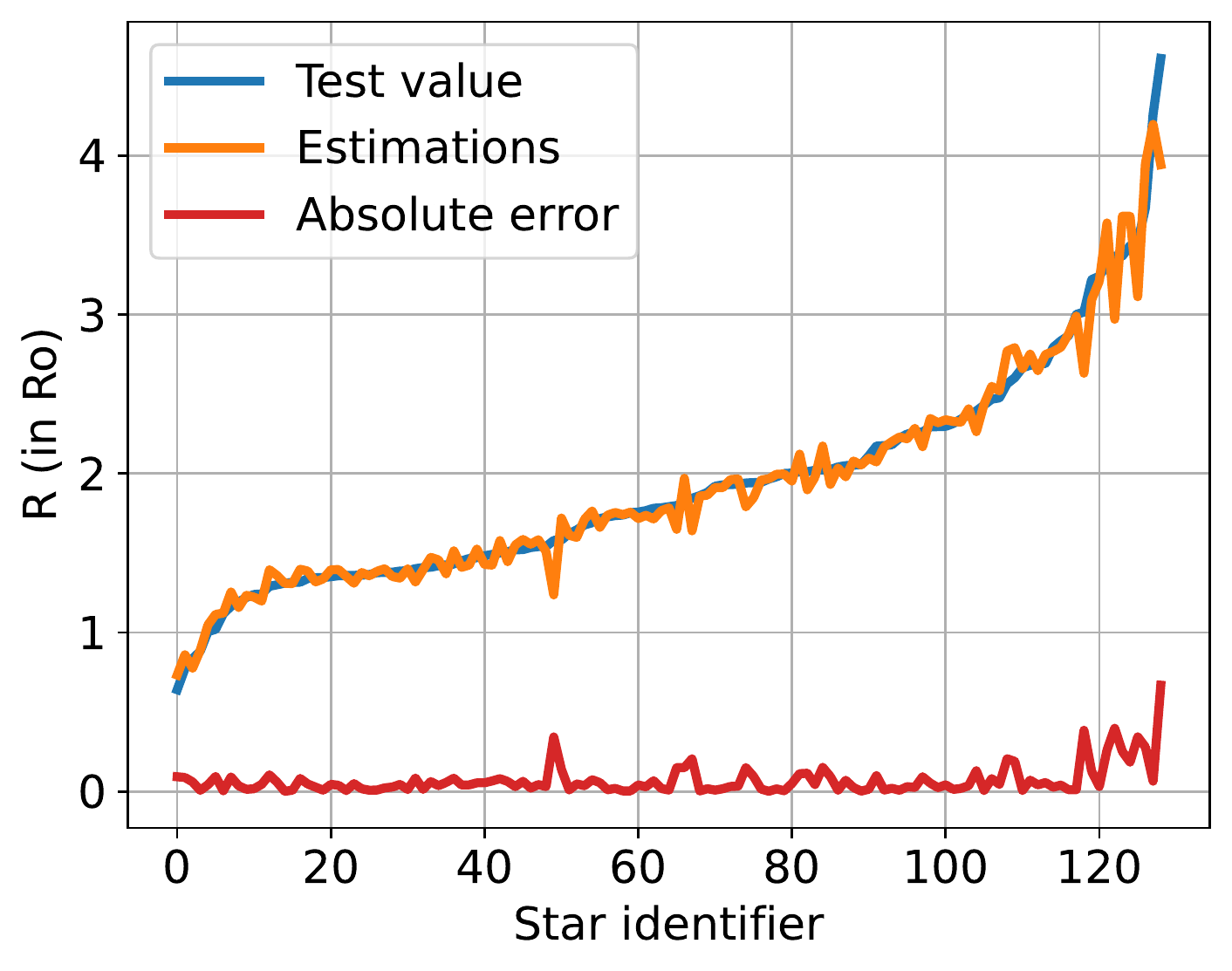}
        \caption{RF}
        \label{fig:ai_models_for_R:rf}
    \end{subfigure}
    %%%%%%%%%%%%%%%%%%%%%%%%%%%%%%%%%%%%%%%%%%%%%%%%%%%%%%%
    \hfill %%%%%%%%%%%%%%%%%%%%%%%%%%%%%%%%%%%%%%%%%%%%%%%%
    %%%%%%%%%%%%%%%%%%%%%%%%%%%%%%%%%%%%%%%%%%%%%%%%%%%%%%%
    \begin{subfigure}{0.2\linewidth}
        \includegraphics[width=\linewidth]{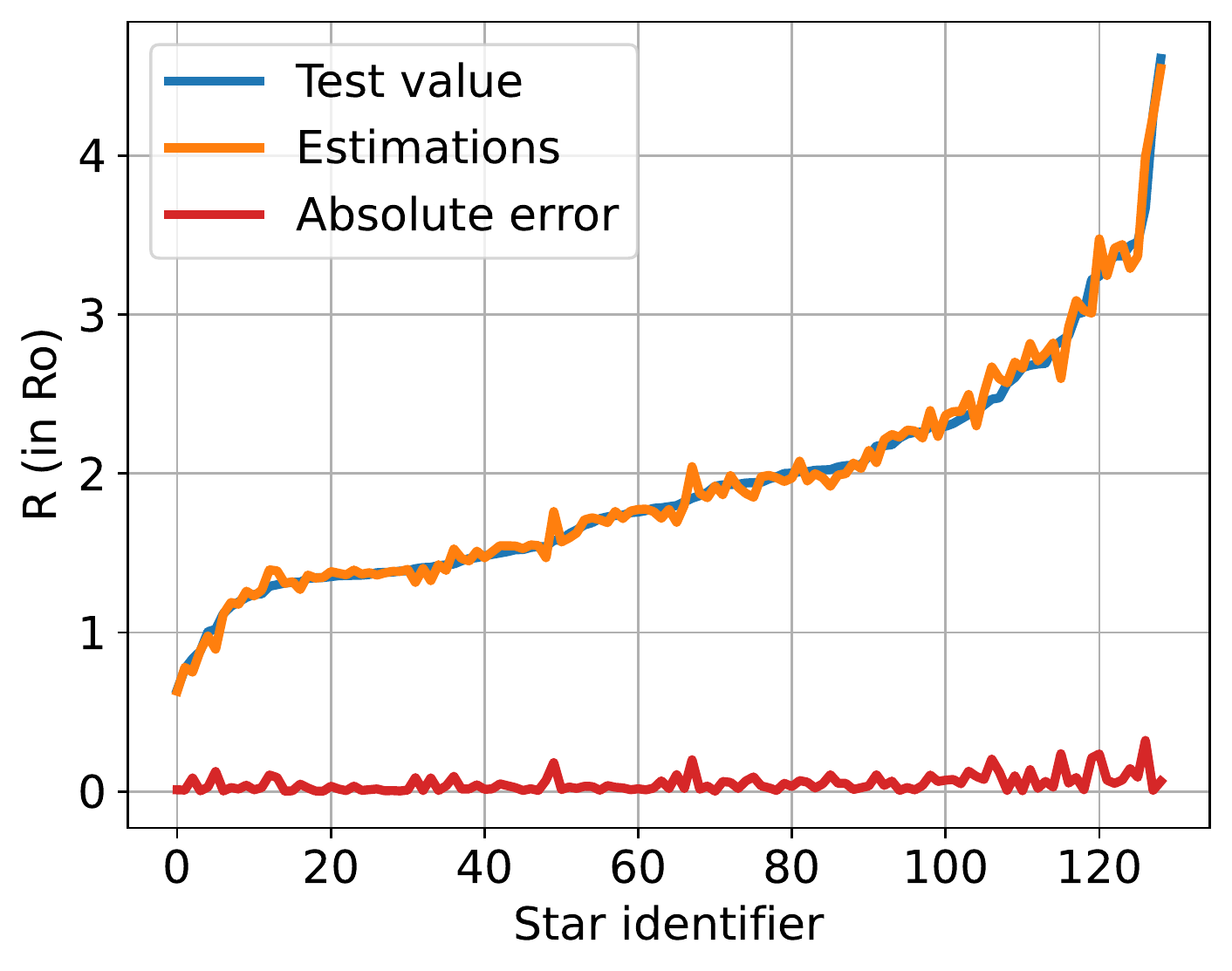}
        \caption{SVR}
        \label{fig:ai_models_for_R:svr}
    \end{subfigure}
    \begin{subfigure}{0.2\linewidth}
        \includegraphics[width=\linewidth]{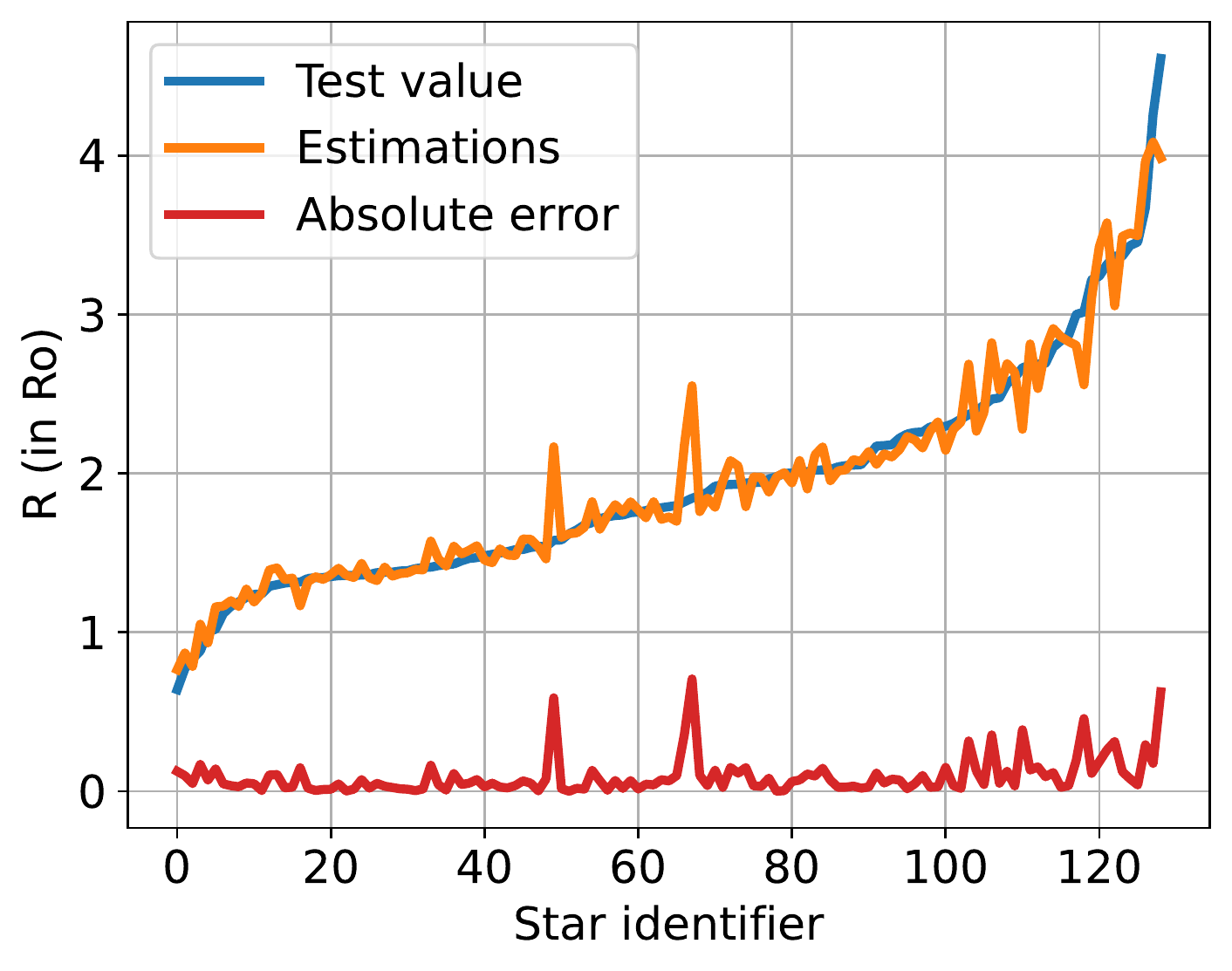}
        \caption{kNN}
        \label{fig:ai_models_for_R:knn}
    \end{subfigure}
    \begin{subfigure}{0.2\linewidth}
        \includegraphics[width=\linewidth]{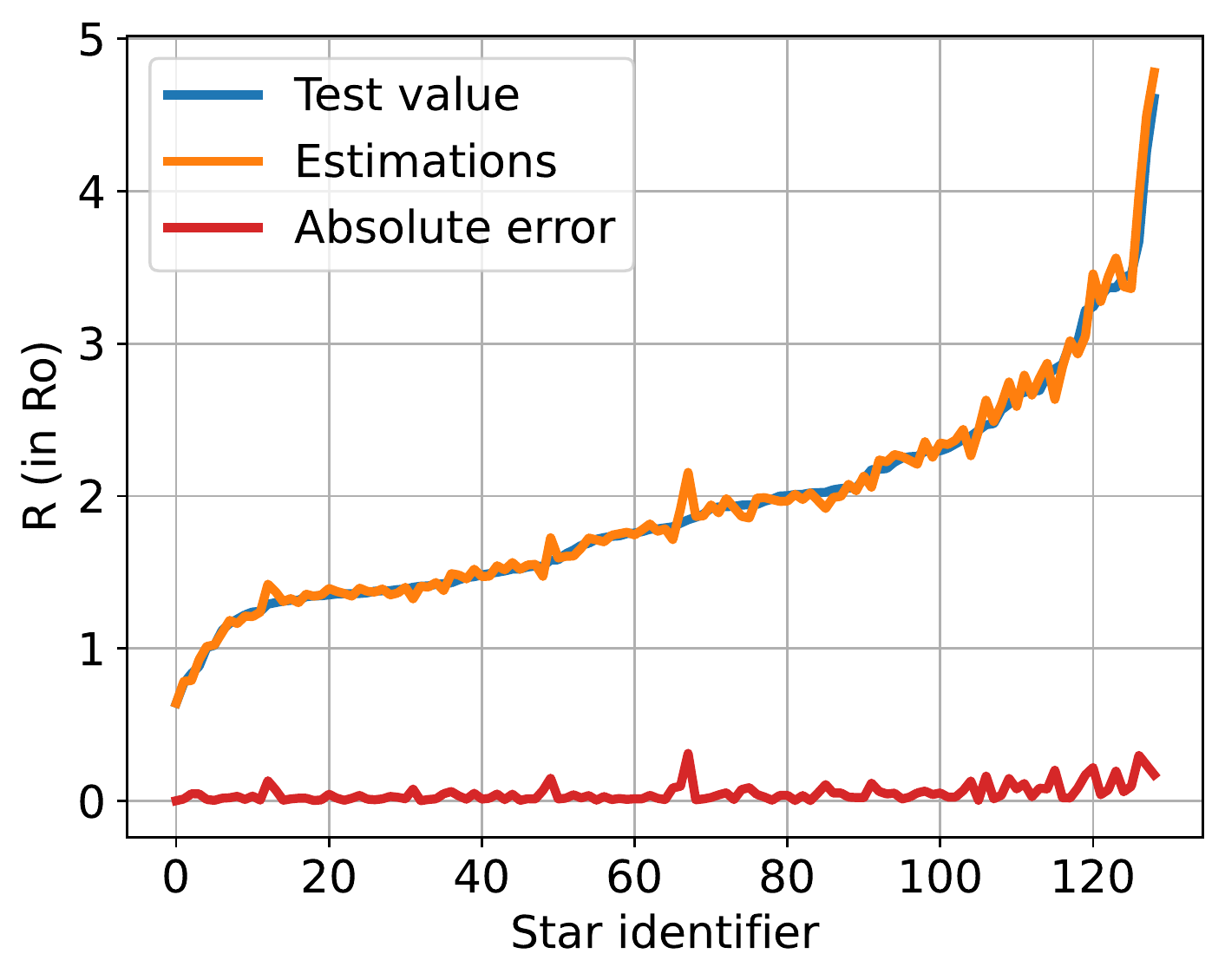}
        \caption{NN}
        \label{fig:ai_models_for_R:nn}
    \end{subfigure}
    \begin{subfigure}{0.2\linewidth}
        \includegraphics[width=\linewidth]{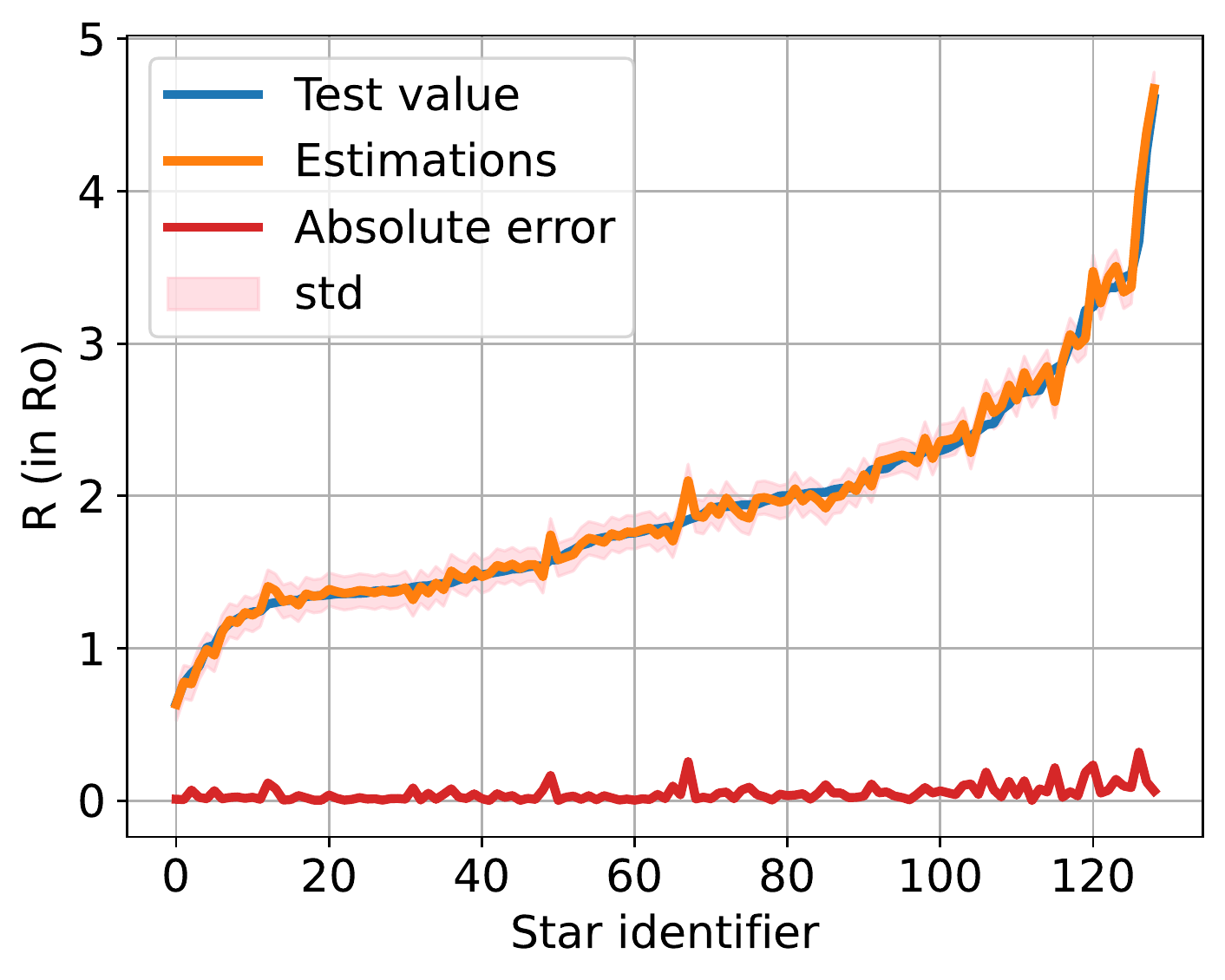}
        \caption{Stacking}
        \label{fig:ai_models_for_R:stacking}
    \end{subfigure}
    \caption{Detailed estimations of the stellar radius for every test sample of all the AI models proposed in this paper. The orange line shows the radii estimated by the AI techniques, and the blue line shows the corresponding test value. The red line represents the absolute error between these two previous quantities. The X-axis shows the identifier of each test star when they have been sorted in increasing order of radius.}
    \label{fig:ai_models_for_R}
\end{figure*}

Figures \ref{fig:ai_models_for_M} and \ref{fig:ai_models_for_R} show the detailed estimations for every test sample for its mass or radius, respectively. One can clearly observe that for the case of the stellar mass, the best estimations are offered by the NN and the Stacking models. For the radius estimations, the winners are the SVR, the NN, and their stacking.

\subsection{Generalization experiment}

The range of masses and radii covered by the data sample is limited and not equally covered by the data sample. For a detailed analysis of the strengths and weaknesses of this data sample we refer the reader to \citet{Moya18}. In summary, there are spectral types statistically well covered by the data sample, especially FGK stars, but beyond theses types the data sample is not that robust. In the near future, large facility surveys and space missions will populate these regions, but in the meantime it is important to analyze whether AI models are able to generalize their estimates to ranges little or not observed during training. Technically, in this second experiment we use the data sample as follows. We first sort in ascending order the star samples in the dataset according to their masses (or radii). Then we train the AI models with the $90\%$ of the stars with the lowest masses (or radii). The test set is conformed with rest of stars ($10\%$), whose masses (or radii) have \emph{not} been observed during training. Specifically, the training and test intervals for the radii are $[0.64, 2.98]$ and $[3.00, 8.35]$, respectively. For the masses, we have the interval $[0.66, 1.49]$ for training, and $[1.50, 2.31]$ is the interval of the test set. Overall, with these splits, we are able to evaluate how the AI models are able to offer estimations for the masses and radii of stars with unobserved characteristics during training.

\begin{table}
\centering
\caption{Generalization experiment. MAE for every AI model.}\label{table:ai_models_generalization}
\begin{tabular}{l|cc}  
\toprule
\textbf{Models} & Mass MAE (in M$_\odot$)& Radius MAE (in R$_\odot$)\\
\midrule
LR & 0.17 & 0.56\\
BRR & 0.17 & 0.56\\
RT & 0.25 & 0.84\\
RF & 0.27 & 0.77 \\
SVR & 0.41 & 0.22\\
kNN & 0.27 & 0.80\\
NN & 0.17 & 0.26 \\
Stacking & \textbf{0.16} & \textbf{0.19}\\
\bottomrule
\end{tabular}
\end{table}

\begin{figure}[ht]
\centering
    \begin{subfigure}{\linewidth}
        \includegraphics[width=0.9\linewidth]{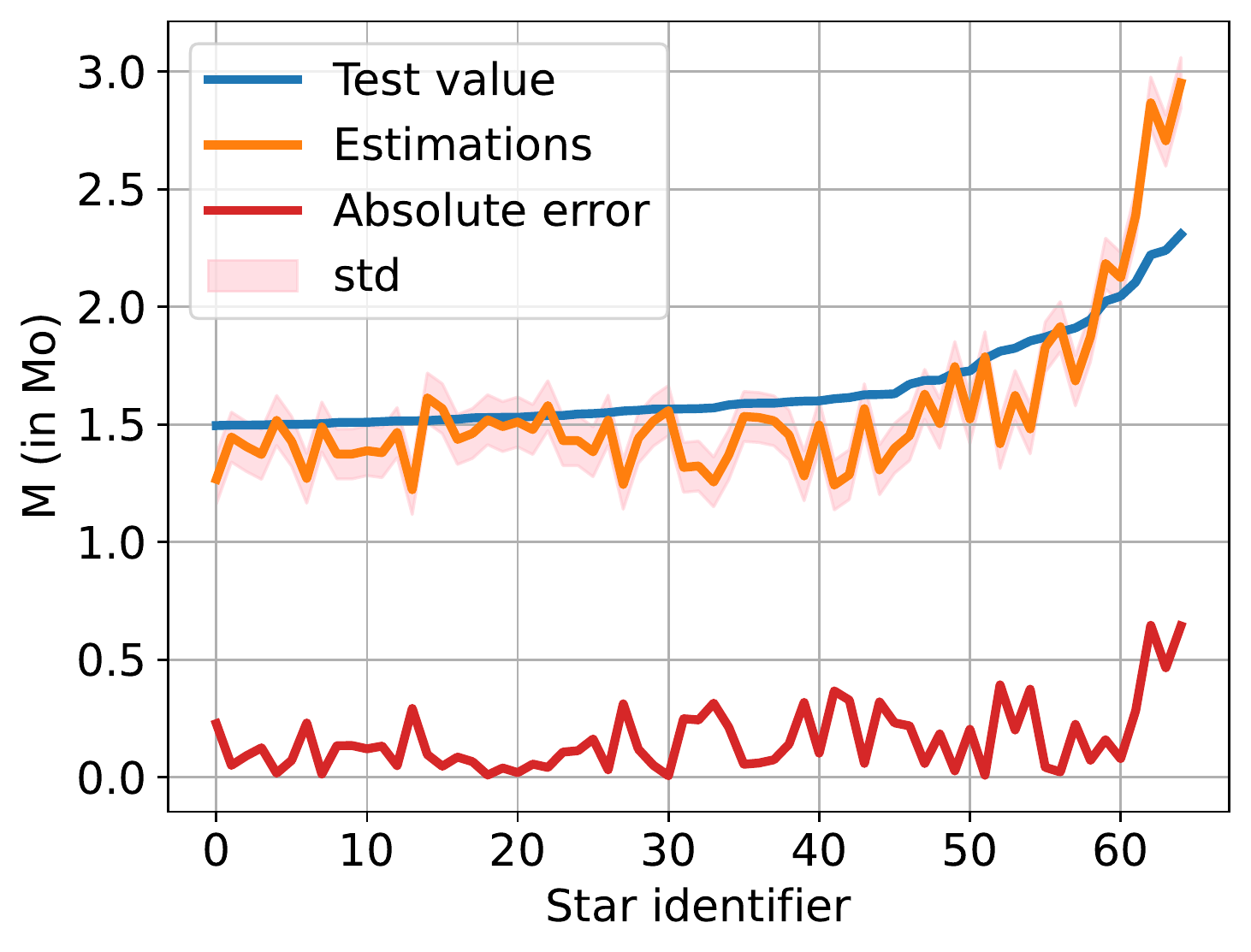}
        \caption{Estimations for the mass.}
    \end{subfigure}
\hfill%\hfil
    \begin{subfigure}{\linewidth}
    \includegraphics[width=0.9\linewidth]{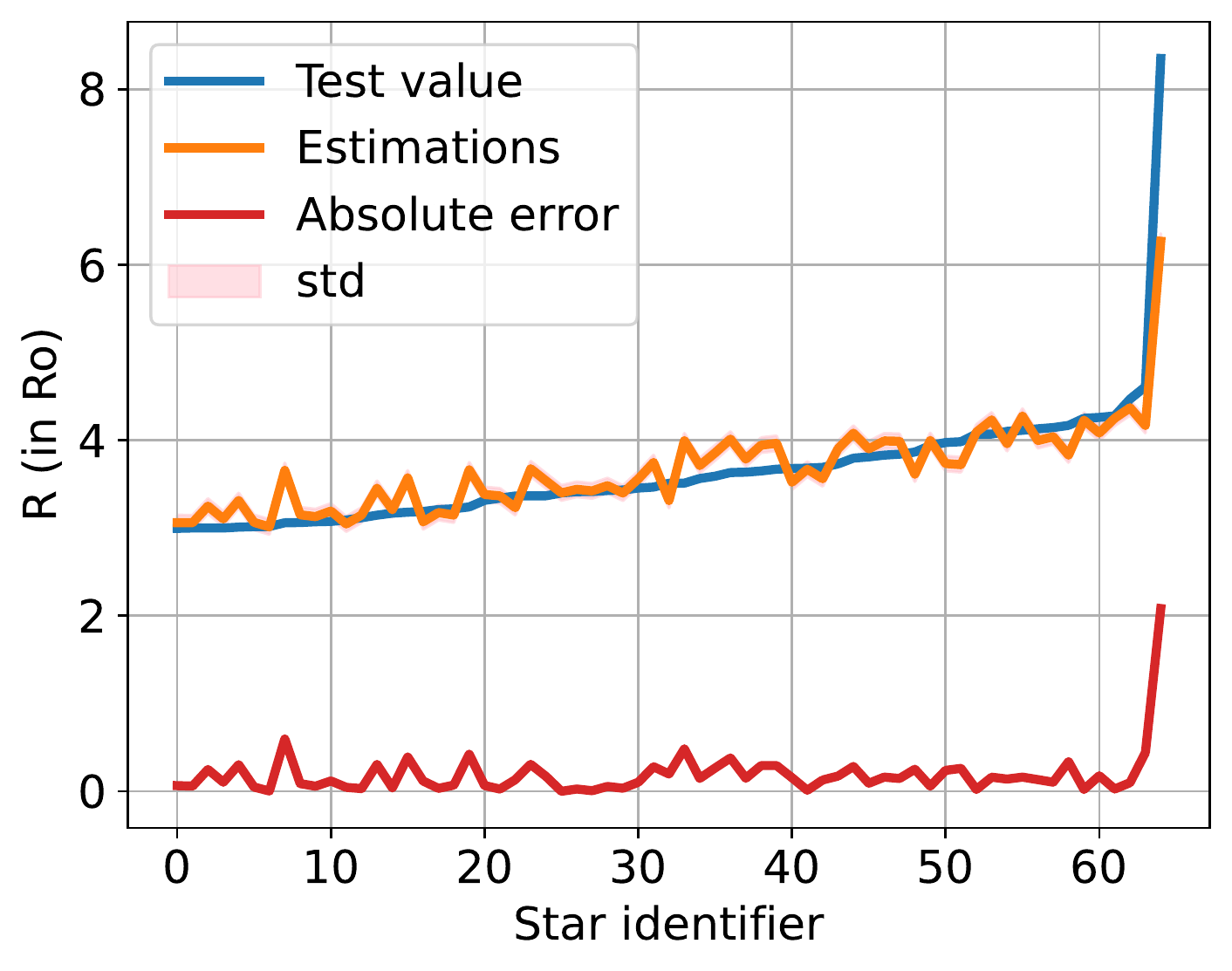}
    \caption{Estimations for the radius.}
\end{subfigure}
    \caption{Generalization experiment. Detailed estimation for the Stacking model. The figure is similar to those of Fig. \ref{fig:ai_models_for_M} and \ref{fig:ai_models_for_R}.}
    \label{fig:generalization_details_stacking}
\end{figure}

Table \ref{table:ai_models_generalization} shows the MAE for all the AI models and every target stellar property. Reader should first note how results have worsened with respect to the previous experiment. This is an indicator of the degree of difficulty offered by the proposed generalization setting. While the previous best result was an MAE of approximately $0.04$, now the errors shift to a higher order of magnitude, being of $0.16$ and $0.19$ for the mass and radius, respectively. Interestingly, generalization results again confirm that an Stacking approach is the best option. This improvement is remarkable in the case of the radius. Our Stacking integrates at level 0 the two best regression models for every target, that is: a) for the mass, we use the LR and the NN; b) for the radius, we employ the SVR and the NN. We show in Figure \ref{fig:generalization_details_stacking} the detailed estimations performed by our Stacking model, where the reader can appreciate how, even for unobserved masses or radii during training, the model shows a tendency to adjust the estimates towards the real value of each sample. For the case of mass, the stacking tends to overestimate values, especially for masses greater than 2 M$_\odot$.  For radius estimations, we observe the excellent generalization capability of the stacking model, which offers accurate predictions up to 6 R$_\odot$. For the last sample of 8.35 R$_\odot$, the model reports an absolute error of $\approx 2$ R$_\odot$.

\subsection{Reducing the number of inputs variables}

\begin{figure*}[ht]
\centering
    \begin{subfigure}{0.4\linewidth}
        \includegraphics[width=\linewidth]{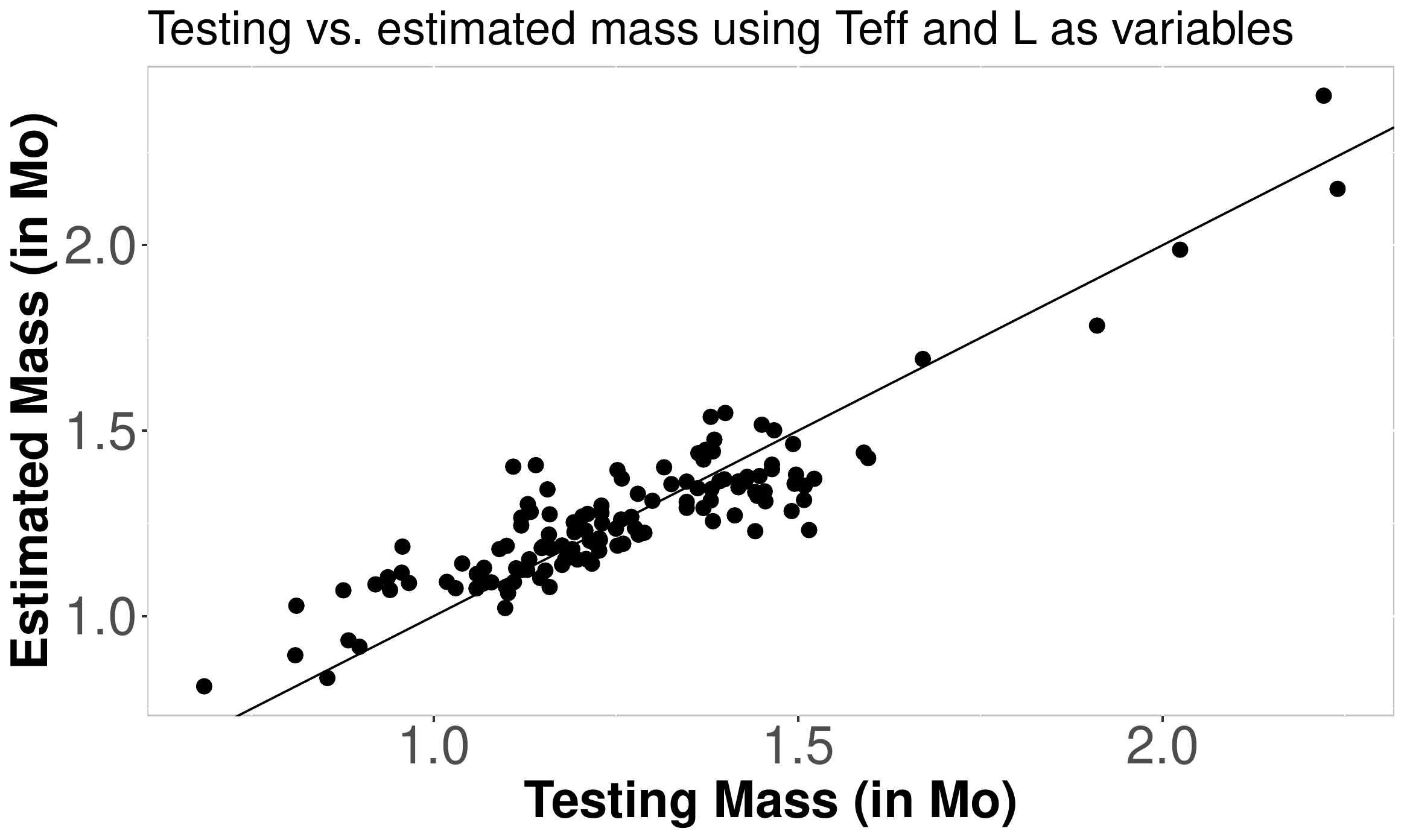}
        \caption{Input: ($T_{\rm eff}$ + $L$).}
        \label{fig:ablation_tl:mass}
    \end{subfigure}
    \begin{subfigure}{0.4\linewidth}
        \includegraphics[width=\linewidth]{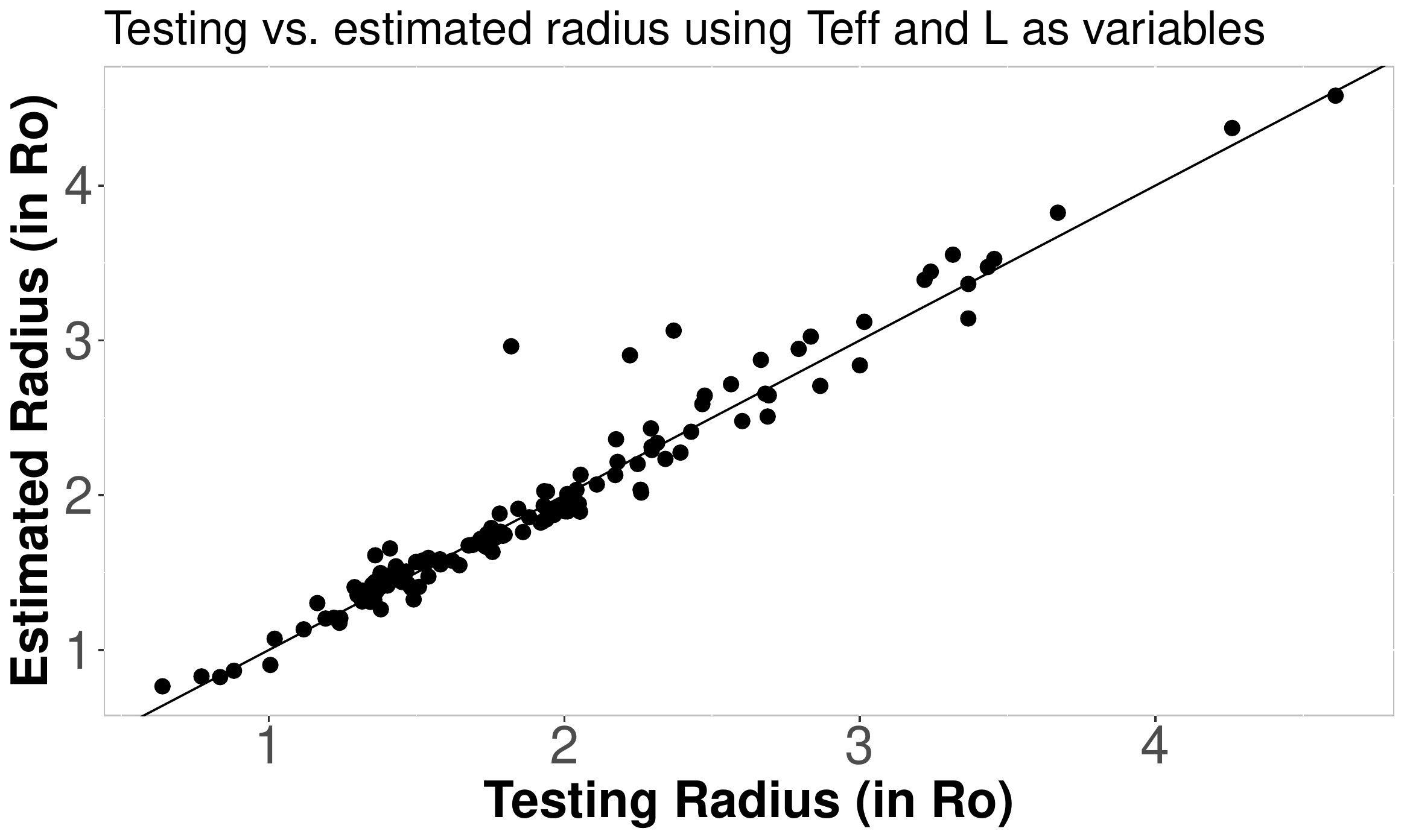}
        \caption{Input: ($T_{\rm eff}$ + $L$).}
    \label{fig:ablation_tl:radius}
\end{subfigure}
\hfill
    \begin{subfigure}{0.4\linewidth}
        \includegraphics[width=\linewidth]{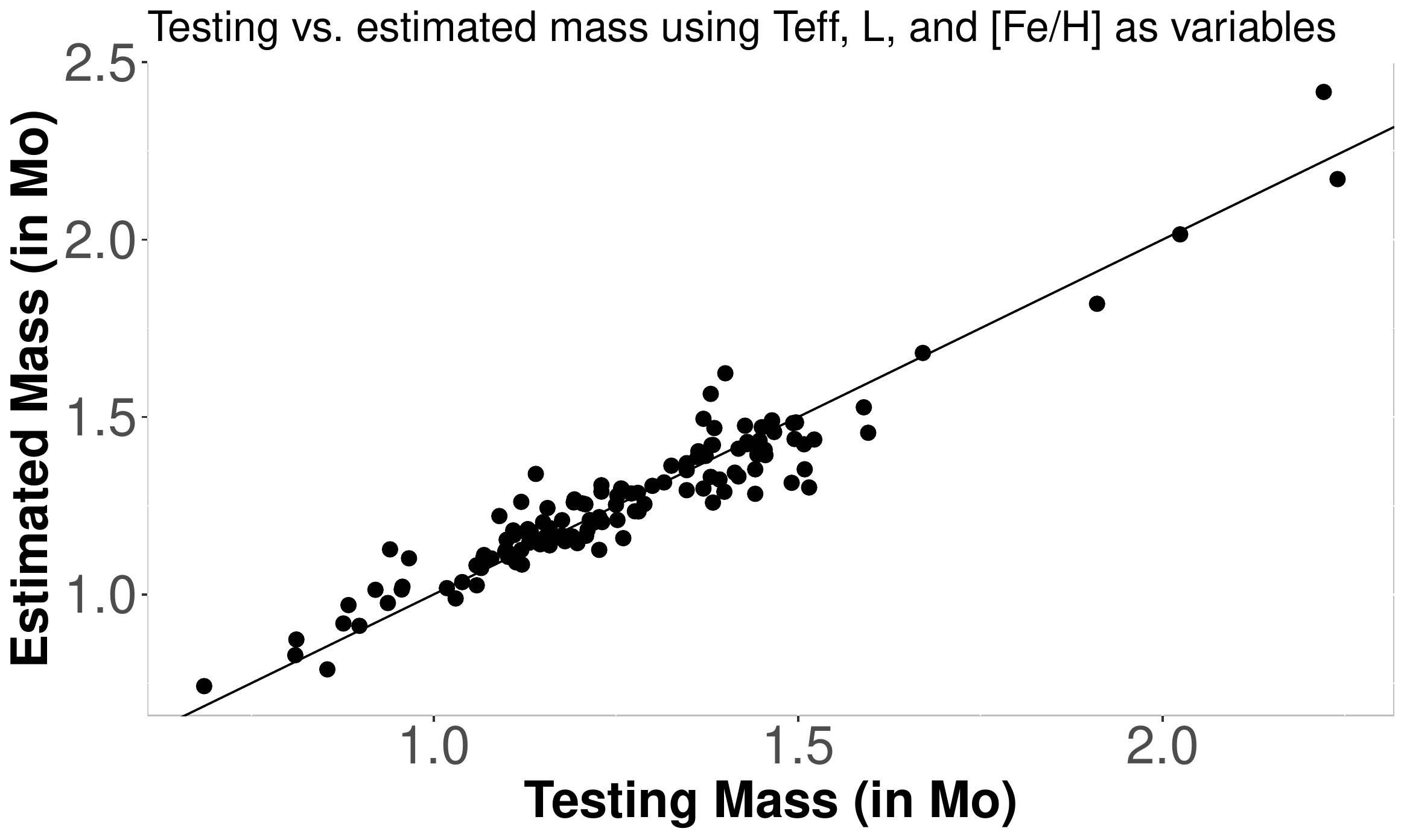}
        \caption{Input: ($T_{\rm eff}$ + $L$ + [Fe/H]).}
        \label{fig:ablation_tlm:mass}
    \end{subfigure}
    \begin{subfigure}{0.4\linewidth}
        \includegraphics[width=\linewidth]{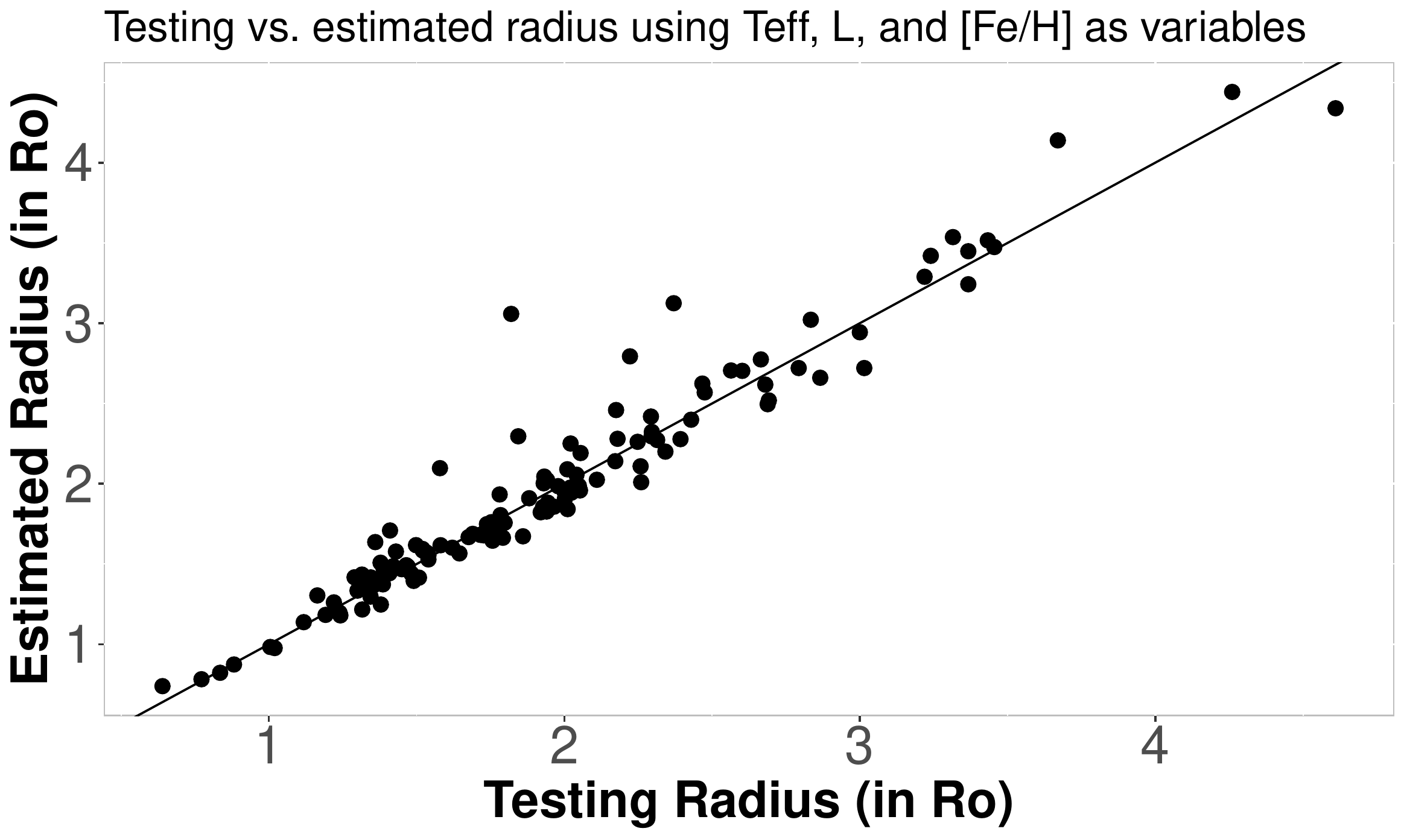}
        \caption{Input: ($T_{\rm eff}$ + $L$ + [Fe/H]).}
        \label{fig:ablation_tlg:radius}
    \end{subfigure}
\hfill
    \begin{subfigure}{0.4\linewidth}
        \includegraphics[width=\linewidth]{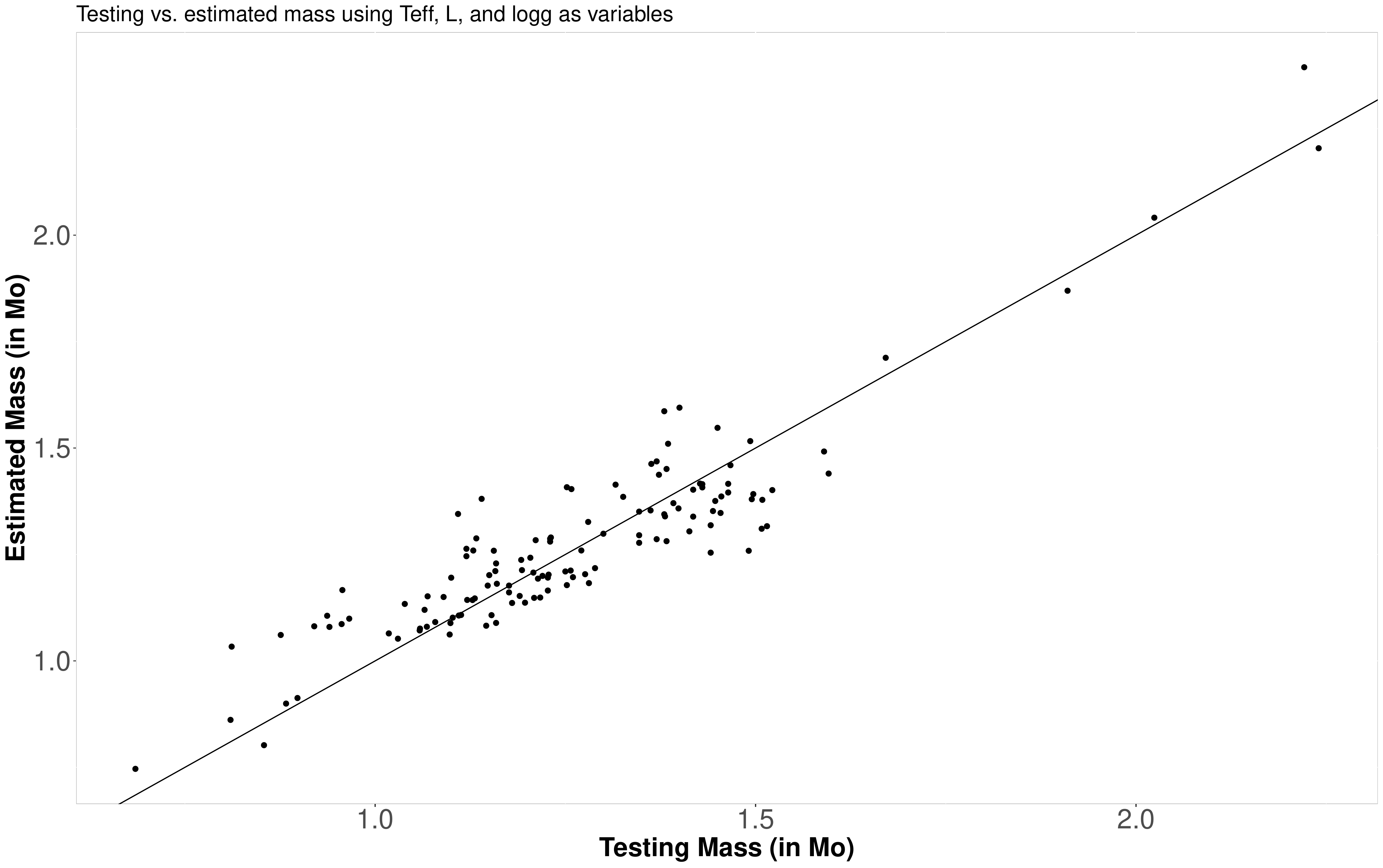}
        \caption{Input: ($T_{\rm eff}$ + $L$ + $\log$g).}
        \label{fig:ablation_tlg:mass}
    \end{subfigure}
    \begin{subfigure}{0.4\linewidth}
        \includegraphics[width=\linewidth]{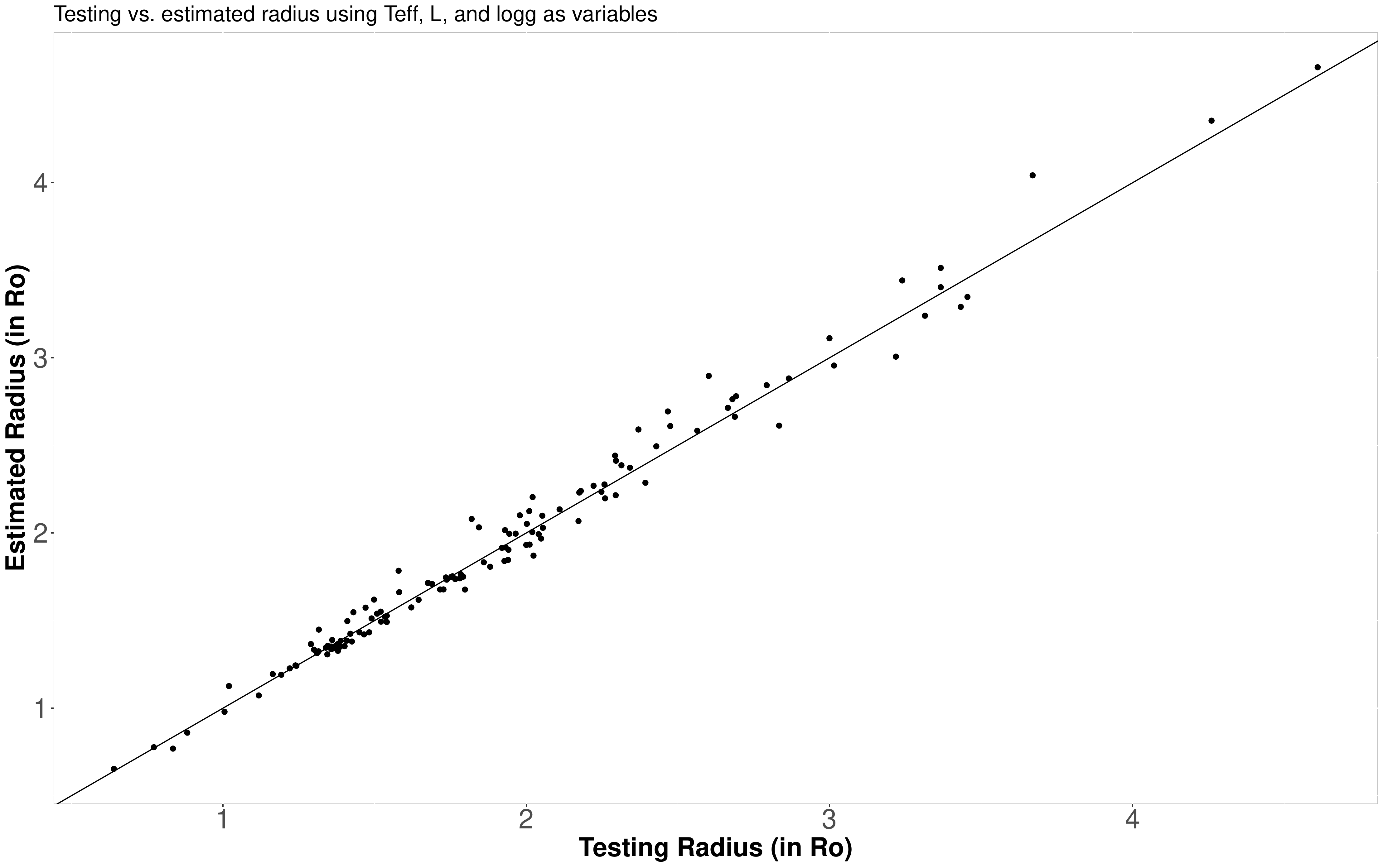}
        \caption{Input: ($T_{\rm eff}$ + $L$ + $\log$g).}
        \label{fig:ablation_tlm:radius}
    \end{subfigure}
\hfill
    \begin{subfigure}{0.4\linewidth}
        \includegraphics[width=\linewidth]{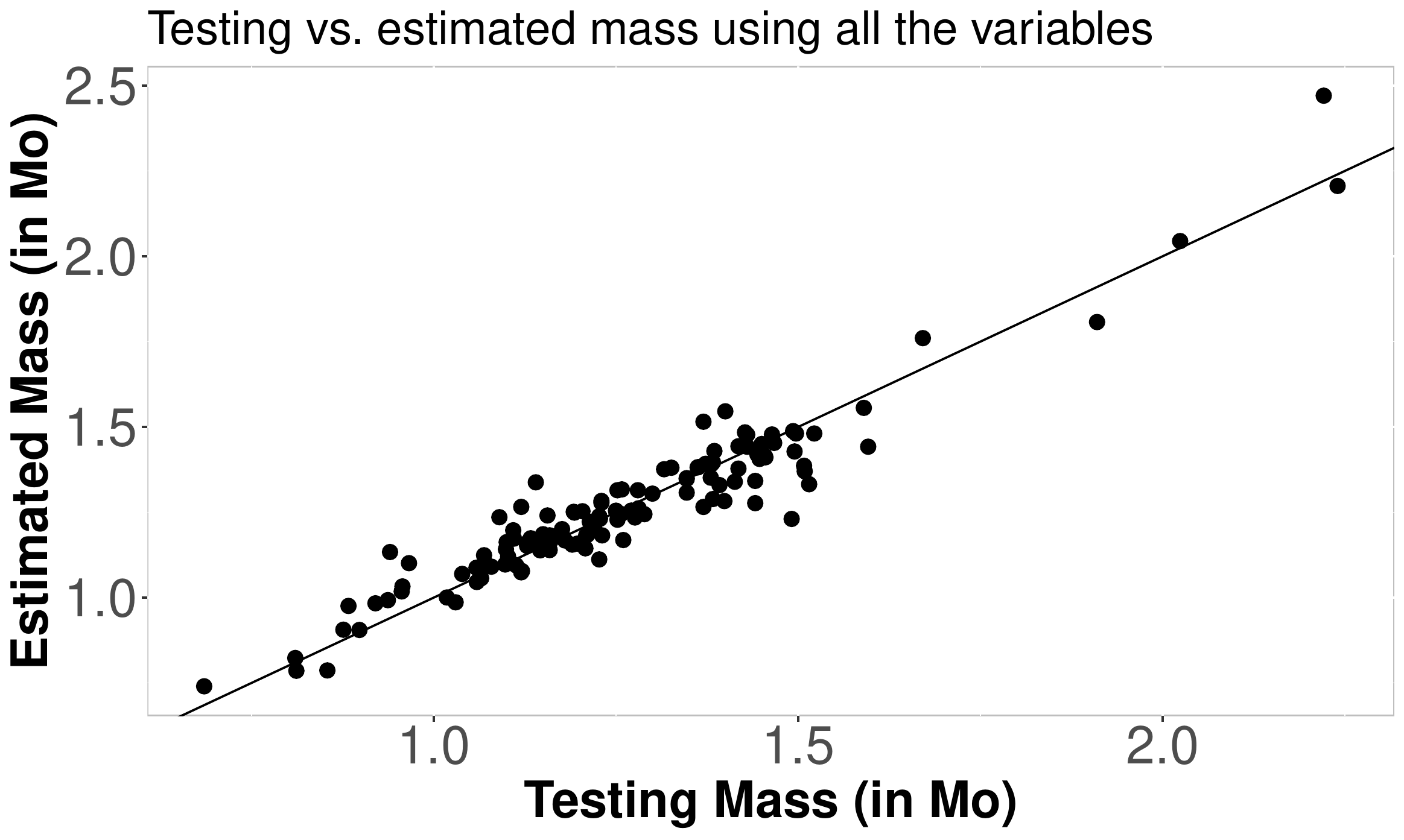}
        \caption{Input: ($T_{\rm eff}$ + $L$ + [Fe/H] + $\log$g).}
        \label{fig:ablation_tlmlogg:mass}
    \end{subfigure}
    \begin{subfigure}{0.4\linewidth}
        \includegraphics[width=\linewidth]{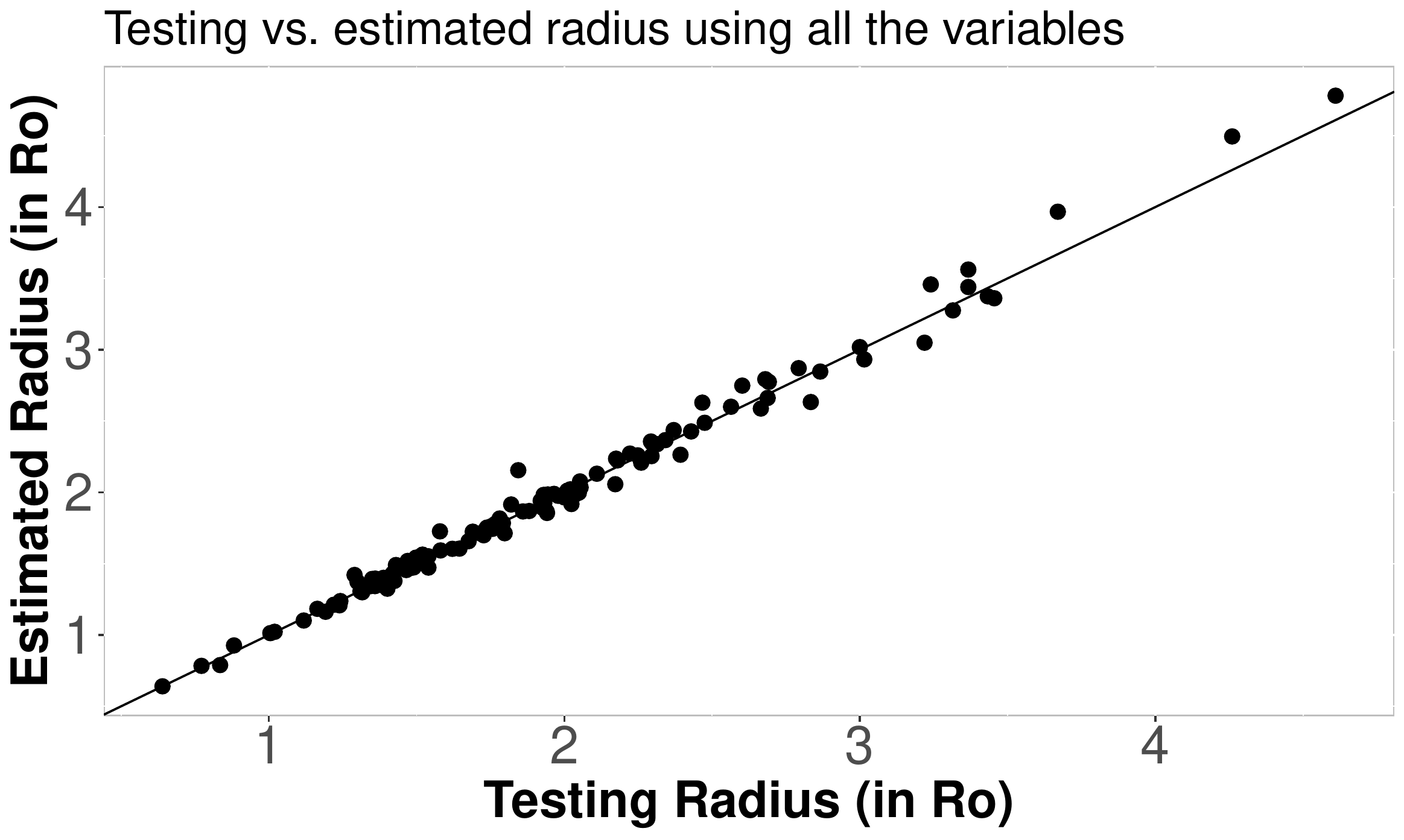}
        \caption{Input: ($T_{\rm eff}$ + $L$ + [Fe/H] + $\log$g).}
        \label{fig:ablation_tlmlogg:radius}
    \end{subfigure}
    \caption{"Testing" vs. estimated masses and radii using the stacking model and the following input features combinations: ($T_{\rm eff}$ + $L$) in a) and b); ($T_{\rm eff}$ + $L$ + [Fe/H]) in c) and d); ($T_{\rm eff}$ + $L$ + $\log$g) in e) and f); ($T_{\rm eff}$ + $L$ + [Fe/H] + $\log$g) in g) and h).}
    \label{fig:ablation}
\end{figure*}

As we mention in Section \ref{sec:sample}, the input independent variables we use for estimating stellar masses and radii are $T_{\rm eff}$, $L$, $\log g$, and [Fe/H]. That is, we assume that the AI model has all this input data at its disposal. Nevertheless, real-life usually is not that good and, for a given star, sometimes we don't have information about all these values. In this section, we seek to measure the importance of each of these input variables with respect to the ability of the model to perform good predictions, and we finally construct models using a reduced set of inputs. Technically, we use our best machine learning model for the study: the stacking. Given these four input features, we proceed to remove some of them during the AI model learning. We start training the stacking model with the simplest and most common combinations of input features: ($T_{\rm eff}$ + $L$). We then incrementally incorporate different features and measure the impact on the model's performance. Figure \ref{fig:ablation} shows our analysis for all these input features combinations.

Regarding the stellar mass we can observe how the estimations become progressively more accurate, as expected, as more features are incorporated into the model: see Figures \ref{fig:ablation_tl:mass}, \ref{fig:ablation_tlm:mass}, \ref{fig:ablation_tlg:mass} and \ref{fig:ablation_tlmlogg:mass}. This effect is even more remarkable in the case of stellar radius, shown in Figures \ref{fig:ablation_tl:radius}, \ref{fig:ablation_tlm:radius}, \ref{fig:ablation_tlg:radius} and \ref{fig:ablation_tlmlogg:radius}. The influence of the logarithm of the surface gravity (\ie $\log g$) is clear for the estimation of the radii, also as expected, since $\log g$ includes explicitly information about $R$.

In Table \ref{table:Dif_inputs} we show the MRD and MARD obtained for estimating $M$ and $R$ on the testing set and the models using $T_{\rm eff}$ + $L$, $T_{\rm eff}$ + $L$ + [Fe/H], $T_{\rm eff}$ + $L$ + $\log g$, and $T_{\rm eff}$ + $L$ + [Fe/H] + $\log g$ as input variables. Here we can see more explicitly what is shown in Fig \ref{fig:ablation}, that is:

\begin{itemize}
    \item The most accurate and unbiased results are obtained when all input variables are used.
    \item For estimating masses, the inclusion of the metallicity remarkably improves the results. $\log g$ is not that important if $T_{\rm eff}$ and $L$ are known. Nevertheless, when only $T_{\rm eff}$ and $L$ are available our model provides a notable accuracy of 6.6 $\%$ and a slight tendency to underestimation.
    \item For estimating radii, it is the inclusion of $\log g$ that significantly improves results. Using only $T_{\rm eff}$ and $L$ provides also remarkable results with an accuracy of 5.3 $\%$ and again a slight tendency to underestimation. The inclusion of metallicity worsens the results. This is an unexpected result but the worsen is small and can be regarded as an effect of increasing variability when adding a new variable without including additional information, that is, according to these results metallicity is not relevant for estimating stellar radii.
\end{itemize}

%Models with a lower number of parameters, or different combinations, can be trained and tested using the codes and data provided in GitHub repository.

\begin{table*}
\centering
%\scalebox{0.9}{
\caption{MRD and MARD using the stacking model and different input variables for estimating the masses and radii of the testing sample.}
\label{table:Dif_inputs}
\begin{tabular}{l|cc|cc|cc|cc}  
\hline
Inputs & \multicolumn{2}{c|}{$T_{\rm eff}$ + $L$ + [Fe/H] + $\log g$} & \multicolumn{2}{c|}{$T_{\rm eff}$ + $L$ + [Fe/H]} & 
\multicolumn{2}{c|}{$T_{\rm eff}$ + $L$ + $\log g$} & 
 \multicolumn{2}{c}{$T_{\rm eff}$ + $L$}\\
\hline
\textbf{Metrics}  & Mass & Radius & Mass & Radius & Mass & Radius & Mass & Radius \\
%\midrule
MRD  & 0.6 & 0.4 & 0.8 & 2.1 & 1.5 & 0.7 & 1.5 & 1.3\\
MARD & 4.1 & 2.3 & 4.3 & 5.8 & 6.1 & 3.2 & 6.6 & 5.1\\
%\bottomrule
\end{tabular}
%}end of scalebox
\end{table*}

\subsection{Comparison with the state of the art}
\label{sec:sota_comparison}

Finally, we proceed to compare our results with the corresponding model based on classical empirical relations between stellar variables published by \citet{Moya18}, considering this work as the state-of-the-art for the problem. In that work, authors compare the performance of their linear regressions with others in the literature, showing that their proposal provides the best balance between accuracy and precision at that time. Technically, for each star in our testing set, we have chosen from \citep{Moya18} the best linear empirical relation for all the features provided in the data sample. This selection is done by finding the empirical relationship with the best combination of accuracy and precision.

\begin{figure}[ht]
\centering
    \begin{subfigure}{\linewidth}
        \includegraphics[width=\linewidth]{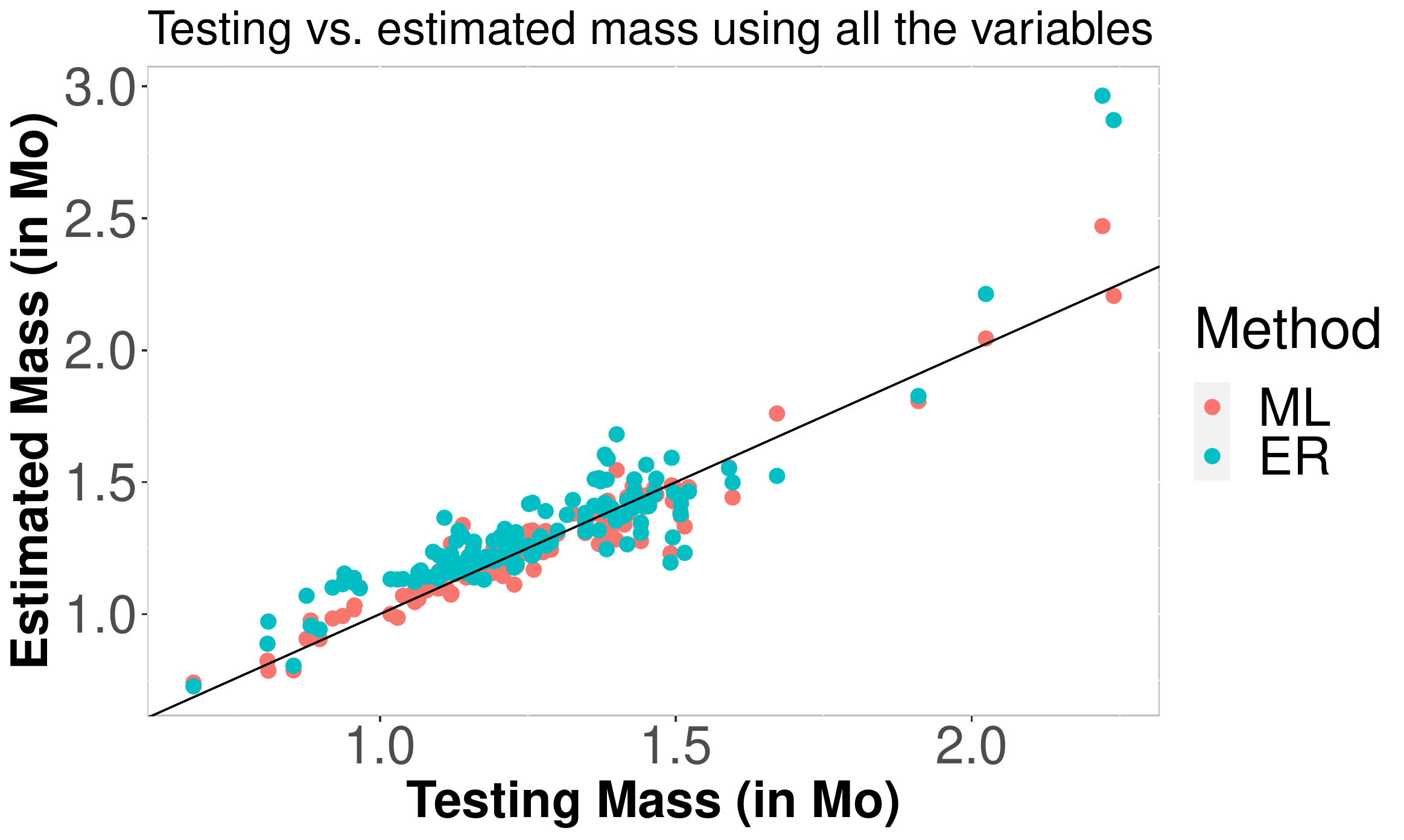}
        \caption{Estimations for the mass.}
        \label{fig:comparison_stacking_vs_er:mass}
    \end{subfigure}
\hfill%\hfil
    \begin{subfigure}{\linewidth}
    \includegraphics[width=\linewidth]{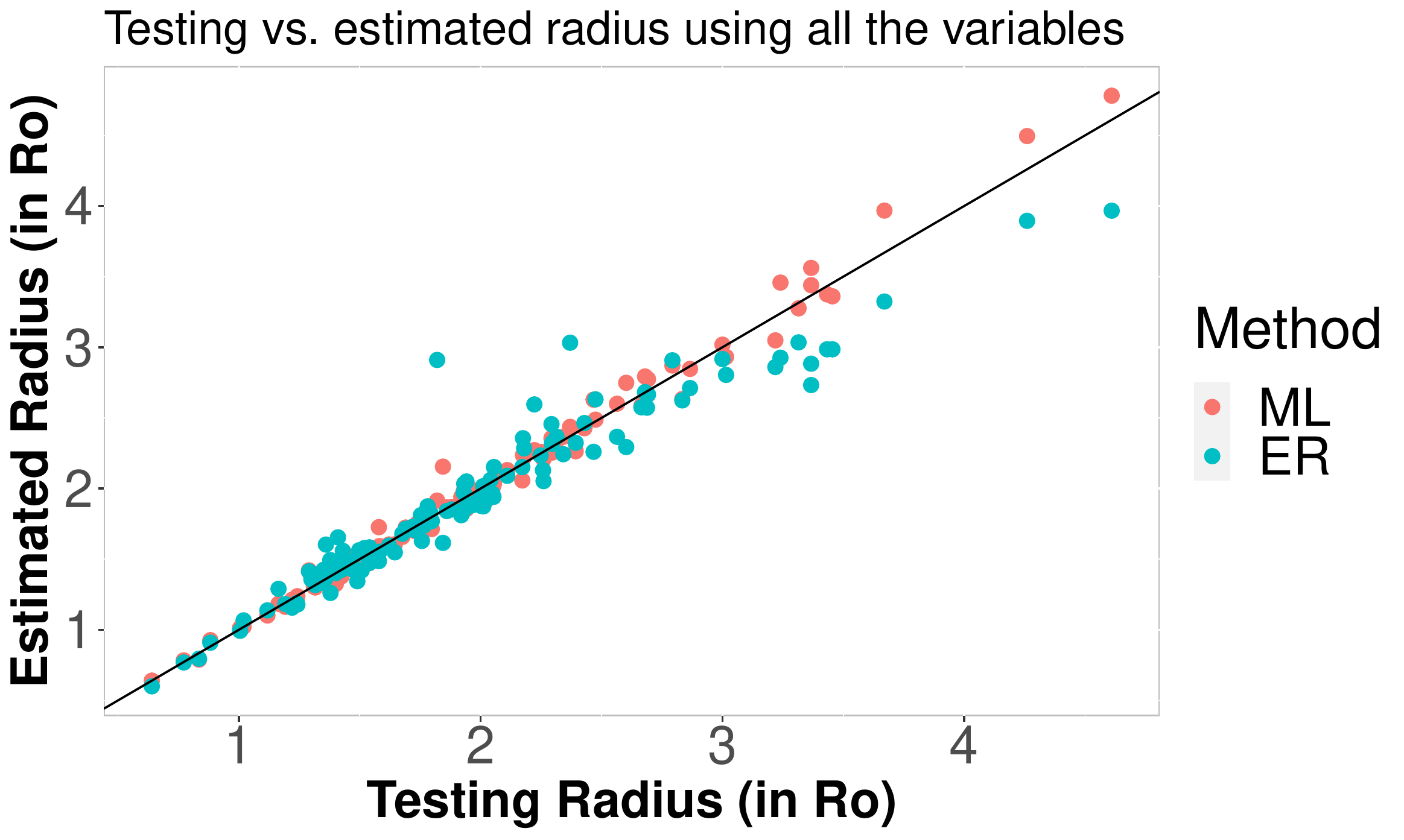}
    \caption{Estimations for the radius.}
    \label{fig:comparison_stacking_vs_er:radius}
\end{subfigure}
    \caption{Comparison of the estimations performed by best machine learning (ML) model, \ie the stacking, and the best empirical relation (ER) in \citep{Moya18}.}
    \label{fig:comparison_stacking_vs_er}
\end{figure}

In Figures \ref{fig:comparison_stacking_vs_er:mass} and \ref{fig:comparison_stacking_vs_er:radius} we overlap the estimations of our best machine learning model, \ie the stacking, and the best empirical relation of \citep{Moya18}, for the mass and radius, respectively. Graphically, the reader can observe that the dispersion for both variables is slightly higher when the empirical relations are used. Table \ref{table:comparison} shows the MRD and MARD metrics for our AI approach and the traditional empirical relations.

\begin{table*}
\centering
%\scalebox{0.9}{
\caption{Comparison between our best AI model, \ie stacking, and the best emprirical relation in \cite{Moya18}.}\label{table:comparison}
\begin{tabular}{l|cc|cc}  
\toprule
 & \multicolumn{4}{c}{\textbf{Methods / Target (Mass or Radius)}}\\
\midrule
\textbf{Metrics}  & AI (Stacking) / Mass & ER \cite{Moya18} / Mass & AI (Stacking) / Radius & ER \cite{Moya18} / Radius \\
\midrule
MRD  & 0.6 & 4.3 & 0.4 & -0.25\\
MARD & 4.1 & 8.4 & 2.3 & 5.3\\
\bottomrule
\end{tabular}
%}%end of scalebox
\end{table*}

For the mass, using the empirical relations we obtain a MARD of 8.4 $\%$ and an MRD of 4.3 $\%$, that is, accuracy better than 10 $\%$ (a result really remarkable for the empirical relations) and a bias of a 4.3 $\%$, lower than the accuracy. However, our AI model based on stacking reduces the bias by a factor of ten and improves the accuracy by a factor of two. Similar results can be observed for the radius. This time the empirical relations report an MARD = 5.3 $\%$ and an MRD = -0.25 $\%$, that is, an accuracy around a 5 $\%$ (a result again remarkable for an empirical relation) and a negligible bias. \citet{Moya18} already highlighted this finding in their previous work, that is, for the radius, the empirical relations provide better results than for the mass. Compared with our AI model, we see that the bias remains negligible and the accuracy is improved again by a factor of two. We can, therefore, conclude that the alternative proposed in this work, using AI models, is able to provide better results for the estimation of stellar masses and radii than any previous empirical relations in \citet{Moya18}. Thus, our work improves the state of the art and provides a clear experimental evaluation, which will allow future comparisons with new methods.

\section{Conclusions}

In this work we have analyzed the effectiveness of different AI techniques for the estimation of stellar masses and radii. For achieving this goal we need a high-quality data sample, where masses and radii have been obtained as accurately as possible. We have used in our experiments the sample in \citet{Moya18}, which is currently the most adequate database for this purpose in the literature. It consists of a total of 726 Main Sequence stars covering spectral types from B to K with the main bulge of them being F-G stars (an 80 $\%$). All of them have precise estimations of $M$, $R$, $T_{\rm eff}$, $L$, $\log g$, and [Fe/H]. 

We have trained and tested a number of AI techniques for estimating masses and radii using the detailed dataset. Specifically, we have evaluated the regression capability of the following AI techniques: Linear Regression, Bayesian Regression, Regression Trees, Random Forest, SVR, kNN, NN, and Stacking. We have designed a series of experiments to study the effectiveness of these techniques for stellar radius and mass estimates. 

We have first divided our data sample into a training set (80$\%$ of the total sample) and a test set (the remaining 20$\%$) and analyzed the accuracy of the AI models with them. This first experiment consists in analyzing the accuracy reached when estimating the testing masses and radii. In both cases, the stacking is the technique showing the best results, with a MAE of 0.049 for the mass, and of 0.048 for the radius. We have found that for estimating stellar masses the NN provides precise results (MAE=0.05), and for estimating stellar radii NN and SVR are two good choices, with a MAE of 0.049 and 0.05 respectively.

In a second experiment, we propose to explore the generalization capacity of these AI techniques. That is, we want to explore the ability of these models to make estimates of radii and masses never seen during training. The interest of this experiment lies in the fact that the data sample covers a certain region of the HR diagram, and we would like to quantify how well masses and radii are estimated in the boundary HR regions defined by the data sample or even beyond them. Our results have confirmed that models based on neural networks and stacking are able to generalize adequately.  In particular, they yield an MAE for mass and radius estimates above the upper bound of the training set of 0.16 and 0.19, respectively.

The simultaneous knowledge of $T_{\rm eff}$, $L$, $\log g$, and [Fe/H] for every target of which we want to estimate its mass and radius is not that common. We have analyzed how the best AI technique (stacking) works when only some of these independent variables are known. We have trained models using only $T_{\rm eff}$ and $L$, and $T_{\rm eff}$, $L$, and [Fe/H], and we have compared them with those estimations obtained using all the variables. For the stellar mass, our results show that when only $T_{\rm eff}$ and $L$ are known, we report an MRD=1.5 and an MARD=6.6, which are remarkable results. The inclusion of [Fe/H] improves significantly the estimations, but the addition of $\log g$ is not that important. In the case of the stellar radius, again using only $T_{\rm eff}$ and $L$ provides an MRD=1.3 and an MARD=5.3. In this case, it is the inclusion of $\log g$ the additional variable improving notably the results, whereas [Fe/H] has a null impact on the results.

We finally compare our best AI model results with the corresponding estimations using the empirical relation of \citet{Moya18}. In \citet{Moya18} we can find a complete comparison with other empirical relations in the literature. For estimating stellar masses, stacking improves the results compared with classical linear regressions, reducing one order of magnitude the bias (MRD) and improving by two the accuracy (MARD). In terms of the stellar radius, the bias is almost unaltered, but the accuracy is again improved by the same factor.

We can see how one of the main benefits of using these AI-based techniques is the accuracy that can be reported. On the other hand, the treatment, propagation and estimation of uncertainties are still one of the limitations of some of these AI models. Some of them have been improved in the last years in this sense, but additional efforts are needed. However, we believe that massive statistical studies, where an accurate estimate of the mass and radius of thousands or more stars is needed, will greatly benefit from the methods and tools we propose in this paper.

The data sample and codes to reproduce all the experiments are available in the following repository: \url{https://github.com/gramuah/ai4mr}. Additionally, we offer an online tool for the estimation of stellar masses and radii using our best AI technique (stacking) and the linear regressions of \citet{Moya18} (\url{http://sdc.cab.inta-csic.es/empiricalRelationsMR}). This online facility is also offered as an R package. Information about this can be found at \url{https://thot-stellar-dating.space/}.

%--------------------------------------------------------------------

\begin{acknowledgements}
The authors acknowledge the referee for his/her very useful and constructive comments. AM acknowledges funding support from the European Union's Horizon 2020 research and innovation program under the Marie Sklodowska-Curie grant agreement No 749962 (project THOT), from Grant PID2019-107061GB-C65 funded by MCIN/AEI/10.13039/501100011033, and from Generalitat Valenciana in the frame of the GenT Project CIDEGENT/2020/036. RJLS acknowledges partial funding support from
project AIRPLANE, with reference PID2019-104323RB-C31, of Spain’s Ministry of Science and Innovation. This research has made use of the Spanish Virtual Observatory (http://svo.cab.inta-csic.es) supported from Ministerio de Ciencia e Innovación through grant PID2020-112949GB-I00. In memoriam of Federico Zuccarino.

\end{acknowledgements}


\bibliographystyle{aa}
\bibliography{references}

\begin{thebibliography}{26}
\expandafter\ifx\csname natexlab\endcsname\relax\def\natexlab#1{#1}\fi

\bibitem[{{Benedict} {et~al.}(2016){Benedict}, {Henry}, {Franz}, {McArthur},
  {Wasserman}, {Jao}, {Cargile}, {Dieterich}, {Bradley}, {Nelan}, \&
  {Whipple}}]{Benedict16}
{Benedict}, G.~F., {Henry}, T.~J., {Franz}, O.~G., {et~al.} 2016, \aj, 152, 141

\bibitem[{Bishop(2006)}]{Bishop06}
Bishop, C.~M. 2006, Pattern Recognition and Machine Learning (Information
  Science and Statistics) (Berlin, Heidelberg: Springer-Verlag)

\bibitem[{Boser {et~al.}(1992)Boser, Guyon, \& Vapnik}]{Boser1992}
Boser, B.~E., Guyon, I.~M., \& Vapnik, V.~N. 1992, in Proceedings of the Fifth
  Annual Workshop on Computational Learning Theory, COLT '92 (New York, NY,
  USA: Association for Computing Machinery), 144–152

\bibitem[{Breiman(2001)}]{Breiman2001}
Breiman, L. 2001, Machine Learning, 45, 5

\bibitem[{Breiman {et~al.}(1984)Breiman, Friedman, Olshen, \&
  Stone}]{Breiman1984}
Breiman, L., Friedman, J.~H., Olshen, R.~A., \& Stone, C.~J. 1984,
  Classification and Regression Trees (Monterey, CA: Wadsworth and Brooks)

\bibitem[{Chang \& Lin(2011)}]{LIBSVM}
Chang, C.-C. \& Lin, C.-J. 2011, ACM Trans. Intell. Syst. Technol., 2

\bibitem[{Drucker {et~al.}(1996)Drucker, Burges, Kaufman, Smola, \&
  Vapnik}]{Drucker1996}
Drucker, H., Burges, C. J.~C., Kaufman, L., Smola, A., \& Vapnik, V. 1996, in
  Proceedings of the 9th International Conference on Neural Information
  Processing Systems, NIPS'96 (Cambridge, MA, USA: MIT Press), 155–161

\bibitem[{Eddington(1926)}]{Eddington}
Eddington, A.~S. 1926, The internal constitution of the stars / by A.S.
  Eddington (University Press Cambridge), viii, 407 p.

\bibitem[{{Eker} {et~al.}(2018){Eker}, {Bak{\i}{\c{s}}}, {Bilir}, {Soydugan},
  {Steer}, {Soydugan}, {Bak{\i}{\c{s}}}, {Ali{\c{c}}avu{\c{s}}}, {Aslan}, \&
  {Alpsoy}}]{Eker18}
{Eker}, Z., {Bak{\i}{\c{s}}}, V., {Bilir}, S., {et~al.} 2018, \mnras, 479, 5491

\bibitem[{{Eker} {et~al.}(2021){Eker}, {Soydugan}, {Bilir}, \&
  {Bak{\i}{\c{s}}}}]{Eker21}
{Eker}, Z., {Soydugan}, F., {Bilir}, S., \& {Bak{\i}{\c{s}}}, V. 2021, \mnras,
  507, 3583

\bibitem[{{Eker} {et~al.}(2015){Eker}, {Soydugan}, {Soydugan}, {Bilir}, {Yaz
  G{\"o}k{\c{c}}e}, {Steer}, {T{\"u}ys{\"u}z}, {{\c{S}}eny{\"u}z}, \&
  {Demircan}}]{Eker15}
{Eker}, Z., {Soydugan}, F., {Soydugan}, E., {et~al.} 2015, \aj, 149, 131

\bibitem[{{Fernandes} {et~al.}(2021){Fernandes}, {Gafeira}, \&
  {Andersen}}]{Fernandes21}
{Fernandes}, J., {Gafeira}, R., \& {Andersen}, J. 2021, \aap, 647, A90

\bibitem[{{Gafeira} {et~al.}(2012){Gafeira}, {Patacas}, \&
  {Fernandes}}]{Gafeira12}
{Gafeira}, R., {Patacas}, C., \& {Fernandes}, J. 2012, \apss, 341, 405

\bibitem[{Goodfellow {et~al.}(2016)Goodfellow, Bengio, \&
  Courville}]{Goodfellow2016}
Goodfellow, I., Bengio, Y., \& Courville, A. 2016, Deep Learning (MIT Press),
  \url{http://www.deeplearningbook.org}

\bibitem[{Hastie {et~al.}(2009)Hastie, Tibshirani, \& Friedman}]{Hastie2009}
Hastie, T., Tibshirani, R., \& Friedman, J. 2009, The Elements of Statistical
  Learning (Springer-Verlag)

\bibitem[{{Hertzsprung}(1923)}]{Hertzsprung}
{Hertzsprung}, E. 1923, \bain, 2, 15

\bibitem[{LeCun {et~al.}(2012)LeCun, Bottou, Orr, \& M{\"u}ller}]{LeCun2012}
LeCun, Y., Bottou, L., Orr, G., \& M{\"u}ller, K.-R. 2012, Efficient BackProp
  (Springer Berlin Heidelberg), 9--48

\bibitem[{{Mann} {et~al.}(2019){Mann}, {Dupuy}, {Kraus}, {Gaidos}, {Ansdell},
  {Ireland}, {Rizzuto}, {Hung}, {Dittmann}, {Factor}, {Feiden}, {Martinez},
  {Ru{\'\i}z-Rodr{\'\i}guez}, \& {Thao}}]{Mann19}
{Mann}, A.~W., {Dupuy}, T., {Kraus}, A.~L., {et~al.} 2019, \apj, 871, 63

\bibitem[{{Moya} {et~al.}(2018){Moya}, {Zuccarino}, {Chaplin}, \&
  {Davies}}]{Moya18}
{Moya}, A., {Zuccarino}, F., {Chaplin}, W.~J., \& {Davies}, G.~R. 2018, \apjs,
  237, 21

\bibitem[{Pedregosa {et~al.}(2011)Pedregosa, Varoquaux, Gramfort, Michel,
  Thirion, Grisel, Blondel, Prettenhofer, Weiss, Dubourg, Vanderplas, Passos,
  Cournapeau, Brucher, Perrot, \& Duchesnay}]{scikit-learn}
Pedregosa, F., Varoquaux, G., Gramfort, A., {et~al.} 2011, Journal of Machine
  Learning Research, 12, 2825

\bibitem[{Robbins \& Monro(1951)}]{sgd}
Robbins, H. \& Monro, S. 1951, The Annals of Mathematical Statistics, 22, 400

\bibitem[{{Russell} {et~al.}(1923){Russell}, {Adams}, \& {Joy}}]{Russell}
{Russell}, H.~N., {Adams}, W.~S., \& {Joy}, A.~H. 1923, \pasp, 35, 189

\bibitem[{{Serenelli} {et~al.}(2021){Serenelli}, {Weiss}, {Aerts}, {Angelou},
  {Baroch}, {Bastian}, {Beck}, {Bergemann}, {Bestenlehner}, {Czekala},
  {Elias-Rosa}, {Escorza}, {Van Eylen}, {Feuillet}, {Gandolfi}, {Gieles},
  {Girardi}, {Lebreton}, {Lodieu}, {Martig}, {Miller Bertolami}, {Mombarg},
  {Morales}, {Moya}, {Nsamba}, {Pavlovski}, {Pedersen}, {Ribas}, {Schneider},
  {Silva Aguirre}, {Stassun}, {Tolstoy}, {Tremblay}, \& {Zwintz}}]{Aldo21}
{Serenelli}, A., {Weiss}, A., {Aerts}, C., {et~al.} 2021, \aapr, 29, 4

\bibitem[{Tipping(2001)}]{Tipping01}
Tipping, M.~E. 2001, J. Mach. Learn. Res., 1, 211–244

\bibitem[{{Torres} {et~al.}(2010){Torres}, {Andersen}, \&
  {Gim{\'e}nez}}]{Torres10}
{Torres}, G., {Andersen}, J., \& {Gim{\'e}nez}, A. 2010, \aapr, 18, 67

\bibitem[{Wolpert(1992)}]{Wolpert1992}
Wolpert, D.~H. 1992, Neural Networks, 5, 241

\end{thebibliography}
\end{document}